\documentclass{article}

\usepackage{arxiv}

\usepackage[utf8]{inputenc} 
\usepackage[T1]{fontenc}    
\usepackage{hyperref}       
\usepackage{url}            
\usepackage{booktabs}       
\usepackage{amsfonts}       
\usepackage{nicefrac}       
\usepackage{microtype}      
\usepackage{lipsum}		
\usepackage{graphicx}
\usepackage{natbib}
\usepackage{doi}
\usepackage{bm}
\usepackage{amsmath}
\usepackage{amssymb}

\graphicspath{{./Figures/}}

\title{Machine-learning identification of the variability of mean velocity and turbulence intensity for wakes generated by onshore wind turbines: Cluster analysis of wind LiDAR measurements}

\author{Giacomo Valerio Iungo\\
	Wind Fluids and Experiments (WindFluX) Laboratory\\ Mechanical Engineering Department\\
	The University of Texas at Dallas\\
	Richardson, TX 75080,  USA\\
	\texttt{valerio.iungo@utdallas.edu} \\
	\And
	Romit Maulik \\
	Argonne National Laboratory\\
	Lemont, IL-60439, USA\\
	\And
	S.  Ashwin  Renganathan\\
	The University of Utah\\
	Salt Lake City,  UT-84112, USA\\
	\And
	Stefano Letizia\\
		Wind Fluids and Experiments (WindFluX) Laboratory\\ Mechanical Engineering Department\\
	The University of Texas at Dallas\\
	Richardson, TX 75080,  USA\\
}

\renewcommand{\shorttitle}{Variability of mean velocity and turbulence intensity for onshore wind turbine wakes}

\hypersetup{
pdftitle={Variability of mean velocity and turbulence intensity for onshore wind turbine wakes},
pdfauthor={G.V. Iungo et al.},
pdfkeywords={Wind turbine wakes, lidar, clustering, POD},
}

\begin{document}
\maketitle

\begin{abstract}
	Wind turbine wakes are the result of the extraction of kinetic energy from the incoming atmospheric wind exerted from a wind turbine rotor. Therefore, the reduced mean velocity and enhanced turbulence intensity within the wake are affected by the characteristics of the incoming wind, turbine blade aerodynamics, and the turbine control settings. In this work, LiDAR measurements of isolated wakes generated by wind turbines installed at an onshore wind farm are leveraged to characterize the variability of the wake mean velocity and turbulence intensity during typical operations encompassing a breadth of atmospheric stability regimes, levels of power capture, and, in turn, rotor thrust coefficients. For the statistical analysis of the wake velocity fields, the LiDAR measurements are clustered through a k-means algorithm, which enables to identify the most representative realizations of the wind turbine wakes while avoiding the imposition of thresholds for the various wind and turbine parameters, which can be biased by preconceived, and potentially incorrect, notions. Considering the large number of LiDAR samples collected to probe the wake velocity field over the horizontal plane at hub height, the dimensionality of the experimental dataset is reduced by projecting the LiDAR data on an intelligently-truncated basis obtained with the proper orthogonal decomposition (POD). The coefficients of only five physics-informed POD modes, which are considered sufficient to reproduce the observed wake variability, are then injected in the k-means algorithm for clustering the LiDAR dataset. The analysis of the clustered LiDAR data, and the associated SCADA and meteorological data, enables the study of the variability of the wake velocity deficit, wake extent, and wake-added turbulence intensity for different thrust coefficients of the turbine rotor and regimes of atmospheric stability. Furthermore, the cluster analysis of the LiDAR data allows for the identification of systematic operations with a certain yaw misalignment of the turbine rotor with the mean wind direction. 
\end{abstract}


\section{Introduction}
Power generation through a wind turbine is based on the extraction of kinetic energy from the incoming atmospheric wind, which, in turn, leads to a reduced mean wind speed and enhanced turbulence intensity past the turbine rotor, namely the generation of a wind turbine wake \cite{Stevens2017,PorteAgel2019}. 

For wind power plants, the proximity of wind turbines can lead to wake interactions, namely for specific wind directions wakes generated by upstream wind turbines can impact downstream turbine rotors entailing reduced power capture and enhanced fatigue loads for the downstream turbines \cite{Bailey2014}. Previous field studies of onshore wind farms showed power losses between 2\% and 4\% of the expected nominal power capture under convective and stable, respectively, atmospheric stability conditions \cite{ElAsha2017}. In contrast, for offshore wind farms, the typical lower turbulence intensity of the incoming wind leads to larger downstream extent of the wind turbine wakes \cite{Christiansen2005}, and more significant power losses, such as between 10\% and 20\% of the annual energy production (AEP) as estimated for the Horns Rev wind farm \cite{Barthelmie2010}, or up to 28\% of the nominal capacity for the Lillgrund wind farm \cite{Sebastiani2020}.

The near-wake characteristics, e.g. velocity deficit, wake width, and wake-added turbulence intensity, are mainly affected by the incoming wind speed, shear, and the aerodynamic characteristics of the turbine rotor, which can be encompassed by the rotor thrust coefficient. In contrast, in the far wake, i.e. downstream to the location where the maximum velocity deficit is achieved \cite{Vermeulen1980}, the evolution of the wake velocity field is mainly dominated by the surrounding atmospheric turbulence, which determines the intensity of the turbulent momentum fluxes promoting the gradual recovery of the wind field to the incoming wind conditions \cite{IungoJTECH2014,ZhanWE2020,ZhanWES2020}.

Reproducing the breadth of the wake variability occurring during normal operations of utility-scale wind turbines for the various parameters determining wind condition (e.g., hub-height wind speed, shear, and veer), atmospheric stability regime (e.g., turbulence intensity, Richardson number, Obukhov length), and turbine setting (e.g., rotor rotational velocity, blade pitch angle, rotor yaw angle) through numerical models and laboratory experiments is very challenging \cite{Veers2019}. Numerical models, for instance, may struggle to generate realistic atmospheric wind conditions, which has recently motivated research to couple mesoscale and microscale models to reproduce large-scale wind heterogeneity and the corresponding smaller-scale wind turbulence \cite{Santoni2020,Draxl2021}. On the other hand, modeling efficiently the action of the turbine blades on the incoming turbulent wind field for numerical simulations is still an active field of research delivering continuous improvements \cite{PorteAgel2011}. 

Laboratory experiments are also becoming a fruitful resource to investigate wind turbine wakes thanks to new wind turbine models reproducing the aerodynamic forcing induced by turbine rotors \cite{Bastankhah2017,Nanos2021}, and setups at the inlet of a test section to reproduce realistic wind conditions \cite{Neuhaus2020}. Nonetheless, the smaller Reynolds numbers and integral length scales of the incoming flow generated in wind and water tunnels are still important limitations to be considered for characterizing and modeling realistic velocity fields of wind turbine wakes.    

Besides numerical simulations and laboratory experiments of wind turbine wakes, field observations of wakes generated by utility-scale wind turbines are becoming instrumental for learning, in more detail, the complex physical processes connected with wind power generation and, in turn, with the generation of wind turbine wakes. Different remote sensing techniques, such as light detection and ranging (LiDAR) \cite{IungoJTECH2013}, radar \cite{Hirth2015}, unmanned aerial vehicles (UAVs) \cite{Kocer2011}, and even instrumented larger airplanes \cite{Platis2018}, have been producing compelling observations of wind turbine wakes over large volumes including the entire downstream extent of wind turbine wakes, yet ensuring sufficient spatial and temporal resolutions. The advancements in remote sensing for probing wind turbine wakes have involved not only the hardware and the technical aspects of the instrumentation, but also the design and the post-processing of the wind data collected under non-stationary and variable conditions, which are typical for the atmospheric wind field. Recently, a framework for the optimal design of field experiments with scanning instruments and retrieval of wind statistics, which is denoted as LiSBOA, has been proposed to maximize the experimental capabilities of the available remote sensing instrumentation and generate statistically accurate measurements of wind turbine wakes \cite{LiSBOA1,LiSBOA2}. 

In this paper, we aim at characterizing the variability under typical operations of utility-scale onshore wind turbines of the wake mean velocity and added turbulence intensity. Wind LiDAR measurements, together with meteorological data collected from a meteorological (met) tower, supervisory control and data acquisition (SCADA) data are leveraged for this study. Considering the large number of LiDAR samples collected for probing a turbine wake at hub height, the experimental data are first projected onto a suitable basis obtained with the proper orthogonal decomposition (POD), to reduce the dimensionality of the classification problem. Subsequently, the coefficients associated with the selected POD modes, which are considered sufficient to reconstruct the observed wake variability, are utilized for a k-means clustering algorithm to generate subsets of the initial LiDAR dataset by grouping together observations ascribed to analogous atmospheric and operational conditions. Once the clustering of the LiDAR data is performed, effects on the subsets of the associated SCADA and meteorological data are also investigated, together with the mean velocity and turbulence intensity fields associated with the wakes of the various clusters.

The remainder of the paper is organized as follows: in section \ref{sec:Exp}, the experimental dataset is described. Then, the dimensionality of the LiDAR data is reduced by applying POD in section \ref{sec:POD}. Subsequently, in section \ref{sec:KMeans} the coefficients of the selected POD modes are analyzed through the k-means algorithm to generate clusters of the LiDAR wake measurements. The results of the cluster analysis on the ensemble statistics of the LiDAR measurements, SCADA, and meteorological data are then discussed in section \ref{sec:Analysis}. Finally, concluding remarks are reported in section \ref{sec:Concl}.

\section{LiDAR experiment for an onshore wind farm}
\label{sec:Exp} 
\begin{figure}[t!]
  \centering
     \includegraphics[width=\textwidth]{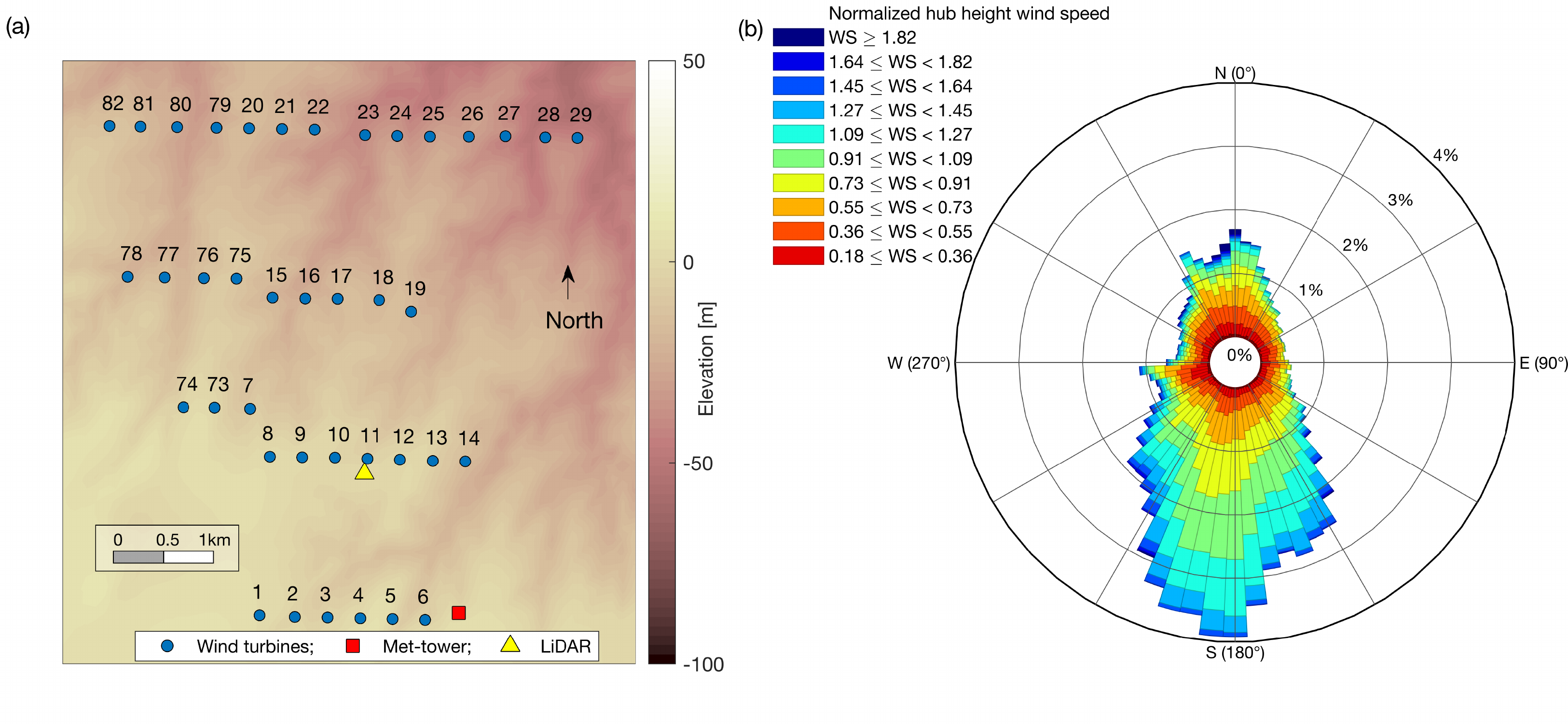}
  \caption{Test site: a) layout of the wind farm, where the size of the blue markers represents down-scaled rotor diameter; b) wind rose of the hub-height wind measured by the met-tower for the entire duration of the experiment and reported as a ratio of the turbine rated wind speed.}
  \label{fig:Map_WD}
\end{figure}

A LiDAR experiment was carried out at a wind farm in North Texas made of 39 Siemens 2.3-MW wind turbines with rotor diameter, $d$, of 108 m, a hub height of 80 m, cut-in wind speed of 3 m/s, rated wind speed of 11.5 m/s, and cut-out wind speed of 25 m/s. The topography map of the site is retrieved from the U.S. Geological Survey \cite{usgs.gov} with a spatial resolution of 100 m (Figure \ref{fig:Map_WD}(a)). By setting the offset altitude at the location of the LiDAR deployment, the standard deviation of the terrain is only 16 m, which allows considering this site as flat terrain. For the retrieval of the LiDAR data, the hub height of each turbine is corrected by taking the local altitude at the turbine locations into account.

The measurement campaign was conducted through various phases between August 2015 and March 2017 for a total of 236 days. Meteorological data were provided as 10-minute averages and standard deviation of wind speed, wind direction, temperature, humidity, and barometric pressure at heights of 36 m, 60 m, 75 m, and 80 m. The atmospheric stability regime is characterized through the Bulk Richardson number\cite{Stull1988}:
\begin{equation}
    \label{eq:Ri}
    Ri_B \left( \overline{z} \right) = \frac{g~\Delta T/\Delta z}{T(z_w)~U^2(z_w)}\overline{z}^2,
\end{equation}
where $g$ is the gravitational acceleration, $z_w=60$ m is the met-tower height where the wind speed, $U$, is measured, $\Delta T$ is the temperature variation over $\Delta z=z_2-z_1$ ($z_2$=75 m and $z_1$=3 m), and $\overline{z}=\sqrt{z_1z_2}$=15 m. 

SCADA data were provided for each turbine as 10-minute averages and standard deviation of wind speed, power output, rotor rotational velocity, and yaw angle. For more details on this dataset and the used quality control process see \citet{ElAsha2017} and \citet{ZhanWE2020}.

The scanning pulsed Doppler wind LiDAR deployed for this experiment is a Windcube 200S manufactured by Leosphere, which emits a laser beam into the atmosphere and measures the radial wind speed, i.e. the velocity component parallel to the laser beam, from the Doppler frequency shift of the back-scattered LiDAR signal. According to the wind farm layout and the prevalence of southerly wind directions (Figure \ref{fig:Map_WD}(b)), for wind directions within the sector between 145$^\circ$ and 235$^\circ$, the wakes produced by the turbines from 1 to 6 evolve roughly towards the LiDAR location, which is a favorable condition for the LiDAR to measure with close approximation the streamwise velocity through plan-position indicator (PPI) scans. Furthermore, according to the layout of Figure \ref{fig:Map_WD}(a), for the considered wind directions, these wind turbines are not affected by upstream wakes.

The LiDAR measurements were typically performed using a range gate of 50 m, an elevation angle of $\phi=3^\circ$, azimuthal range of 20$^\circ$, a rotation speed of the scanning head of 2$^\circ$/s, leading to a typical scanning time for a single PPI of 10 s. After rejecting LiDAR data with a carrier-to-noise ratio (CNR) lower than -25 dB, a proxy for the streamwise velocity is obtained through the streamwise equivalent velocity,  
$U_{eq}=V_r/[cos \phi ~ cos ( \theta - \theta_w ) ]$, where $V_r$ is the LiDAR radial velocity, $\theta$ is the azimuthal angle of the LiDAR laser beam, and $\theta_w$ is the wind direction \cite{ZhanWE2020}. The streamwise equivalent velocity is then made non-dimensional through the velocity profile in the vertical direction of the incoming boundary layer, which is also measured through the wind LiDAR. 

The reference frame used has $x$-direction aligned with the wake direction, which is estimated with the linear fitting of the wake centers at various downstream locations, $y$-direction in the horizontal transverse direction, and $z$-direction vertically and positive moving upwards. The transverse position of the wake center is defined as the location of the minimum velocity obtained by fitting the velocity data at a specific downstream distance through a Gaussian function. More details on the LiDAR system, the field campaign, and data post-processing are available in \citet{ZhanWE2020}.

A total number $N$ = 6,654 quality-controlled PPI LiDAR scans of isolated wind turbine wakes have been processed to provide the non-dimensional wake velocity fields used for this study \citep{ZhanWE2020,ZhanWES2020}. A coordinate transformation is performed to estimate the wake velocity over the horizontal plane at hub height. Specifically, a pseudo radial coordinate, $r$ is defined as:
\begin{equation}
    \label{eq:r}
    r=sign(y)\sqrt{y^2+z^2},
\end{equation}
noting that the $r$ coordinate has sign to identify the two sides from the $x$-axis of the wake velocity field. The wake velocity fields measured with the LiDAR are then interpolated over the horizontal plane at hub height within the domain $1 \leq x/d \leq 7$ and $-1 \leq r/d \leq 1$. The wake velocity data have a spatial resolution of 0.1$d$ and 0.05$d$ in the $x$ and $r$ directions, respectively, generating a data matrix of $[p\times q] =[ 61 \times 41]$ ($pq=2,501$) for a single LiDAR snapshot. 

\begin{figure}[t!]
    \centering
    \includegraphics[width=\textwidth]{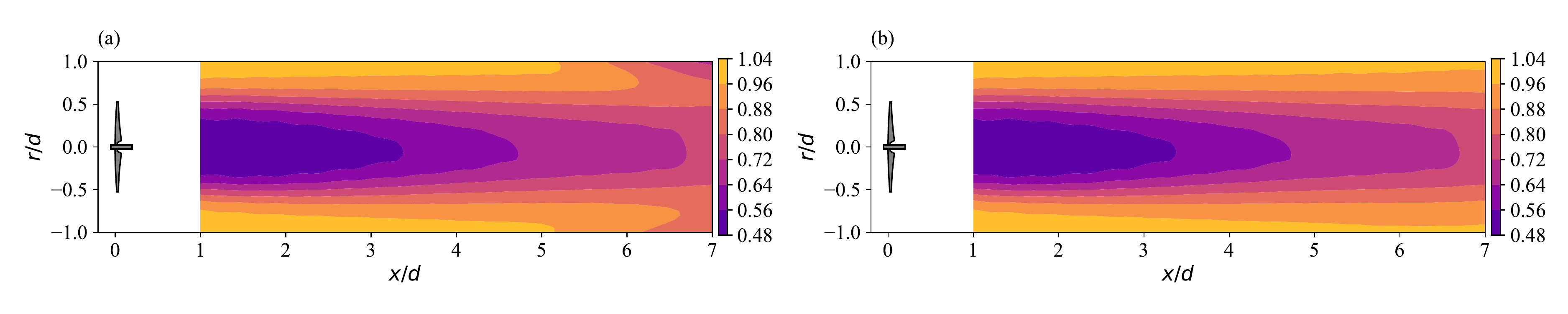}
    \caption{Ensemble mean of the non-dimensional wake velocity fields calculated over the entire dataset: a) NaN values interpolated with the function inpaint-nans; b) interpolation-free data.}
    \label{fig:Mean}
\end{figure}

For the wake analysis only LiDAR samples with $|z/d|<0.25$ are considered to ensure that they are representative of the wake velocity field at hub height. For grid points where the LiDAR data are not available, which can be due to the quality control process of the LiDAR data or a significant misalignment between the wake direction and the direction connecting the LiDAR and turbine locations, the velocity fields are interpolated through the inpaint-nans function available in Matlab \citep{DErrico2004}. The ensemble-averaged velocity fields calculated over the entire dataset are reported in Figure \ref{fig:Mean} for both raw and interpolated data. The main data distortion due to the velocity interpolation occurs at the downstream corners of the spatial domain, which are the areas where the probability of missing LiDAR samples is higher. The interpolated velocity fields are only used for the POD, which does not allow for not-a-number (NaN) values over the spatial domain for the calculation of the eigenproblem of the velocity covariance matrix. In contrast, statistics of the wake velocity field will be calculated with the original non-interpolated velocity fields by ignoring the NaN values.

\section{Proper Orthogonal Decomposition of the LiDAR dataset}
\label{sec:POD}
The analysis of a large dataset encompassing correlated variables aiming to detect trends, data variability, and features can be very challenging and computationally demanding. Specifically, considering a cluster analysis with a total number of $K$ clusters ($K\approx10$ for this work), $I$ iterations for the k-means algorithm ($I\approx10K$), and $L$ repetitions ($L\approx100$), the total number of integrals to be computed is $L\times I \times K \times N \sim \mathcal{O}(10^9)$~~\cite{Li2020}. Therefore, rather than analyzing the LiDAR dataset in its entirety, it can be more convenient to represent the data on a lower-dimensional subspace with a suitable lower rank to enable a simplified analysis, which typically entails lower computational costs and clearer interpretation of the results \cite{Iungo2011a,Iungo2011b,Debnath2017}. Specifically, POD \cite{Lumley1970,berkooz1993proper,Holmes1996} allows for the generation of an orthonormal basis, which is optimal for the reconstruction of the data variability. The most computationally expensive task of POD is the calculation of the correlation matrix of the LiDAR measurements (proportional to $N \times (N+1) / 2$). Therefore, the computational saving is about two orders of magnitude $(N+1)/(2L \times I \times K)=\mathcal{O}(10^{-2})$.

POD is computed for the non-dimensional interpolated LiDAR data with the method of snapshots \cite{Sirovich1987}. A snapshot of the wake velocity field measured with the LiDAR, $u$, can be represented through a linear combination of deterministic functions, which are referred to as POD modes $\phi_j$:
\begin{equation}
    \label{eq:PODreconstr}
    u(x, y, z, t)=\sum_{j=1}^{pq-1} \phi_j (x, y, z)~ a_j(t),
\end{equation}
where $t$ is time. 
In this study, the parameters $a_j(t)$ are coefficients representing the amplitude of each POD mode. POD provides a modal decomposition that is completely a-posteriori and data-dependent, which does not neglect the non-linearities of the original dynamical system, even being a linear procedure. Furthermore, the POD basis is orthonormal and optimal in variance, i.e. among all linear decomposition techniques, it provides the most efficient detection, in a certain least-squares optimal sense, of the dominant components. 

The LiDAR dataset, $\mathbf{U}$, with dimensions $[pq \times N]$ can be approximated by computing the first $r$ most energetic principal components through the singular value decomposition (SVD):
\begin{equation}
    \label{eq:POD}
    \mathbf{U} \approx \Phi \times \Sigma \times V^{\top},
\end{equation}
where $\Phi$ ($[pq \times r]$) and $V$ ($[N \times r]$) are orthonormal matrices and $\Sigma$ is a diagonal matrix ($[r \times r]$) with the first $r$ singular values of $\mathbf{U}$ in descending order as diagonal entries \cite{James_2013B}. Each diagonal element of $\Sigma$, $\sigma_j$, represents the energy contribution of the POD mode $\phi_j$ to the covariance matrix of the velocity snapshots. The POD modes are obtained as columns of the matrix $\Phi$, while the principal components, $a_j$, associated with each POD mode are obtained by projecting the snapshot dataset onto the POD basis $ \mathbf{A} = \mathbf{U}^{\top} \Phi$, where the principal components, $a_j$, are the columns of $\mathbf{A}$, whose size is $[N \times r]$. 

\begin{figure}[h!]
    \centering
    \includegraphics[width=\textwidth]{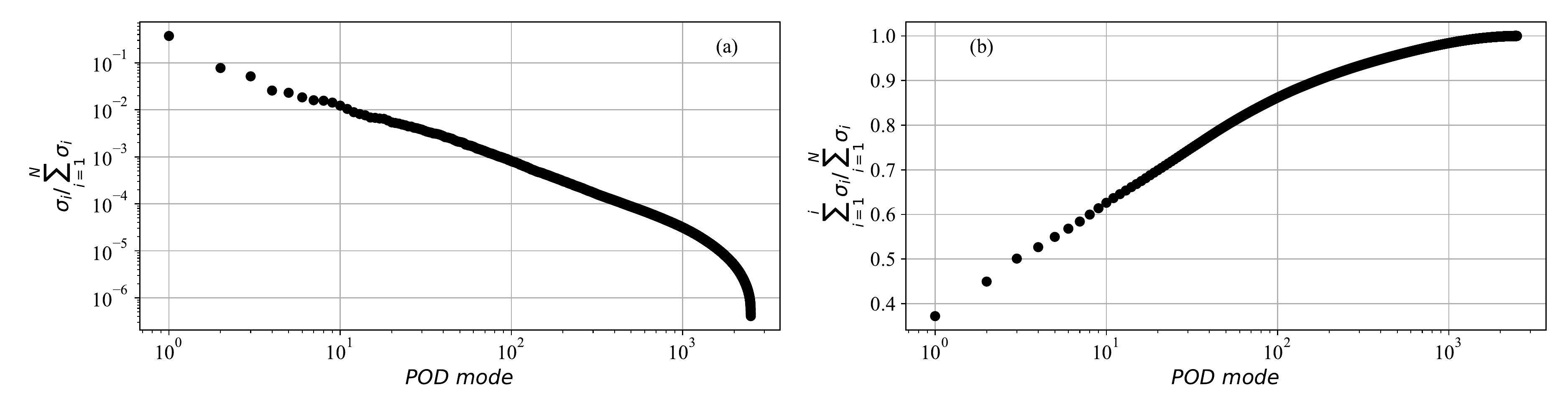}
    \caption{Eigenvalues of the covariance matrix of the LiDAR snapshots, $\Sigma$: a) eigenvalues, $\sigma_i$; b) cumulative POD energy.}
    \label{fig:PODene}
\end{figure}

POD is applied to the non-dimensional interpolated LiDAR wake measurements collected over the horizontal plane at hub height. The obtained eigenvalues of the POD modes, $\sigma_j$, and the respective cumulative energy reconstructed with an increasing number of POD modes are reported in Figure \ref{fig:PODene}. It is now crucial to select the smallest number of POD modes enabling an efficient reconstruction of the wake variability probed through the LiDAR measurements. The selection of POD modes can be done based on the energetic contribution of the various POD modes. In other words, by leveraging the energy optimality of the POD basis, the latter is truncated to reconstruct a certain percentage of the total energy of the velocity covariance matrix (see Figure \ref{fig:PODene}(b)). An alternative to this energetic approach has been used for the present work, which consists of visually inspecting the most energetic POD modes and selecting only POD modes whose spatial morphology indicates a clear physical feature \cite{Debnath2017}.

\begin{figure}[h!]
    \centering
    \includegraphics[width=\textwidth]{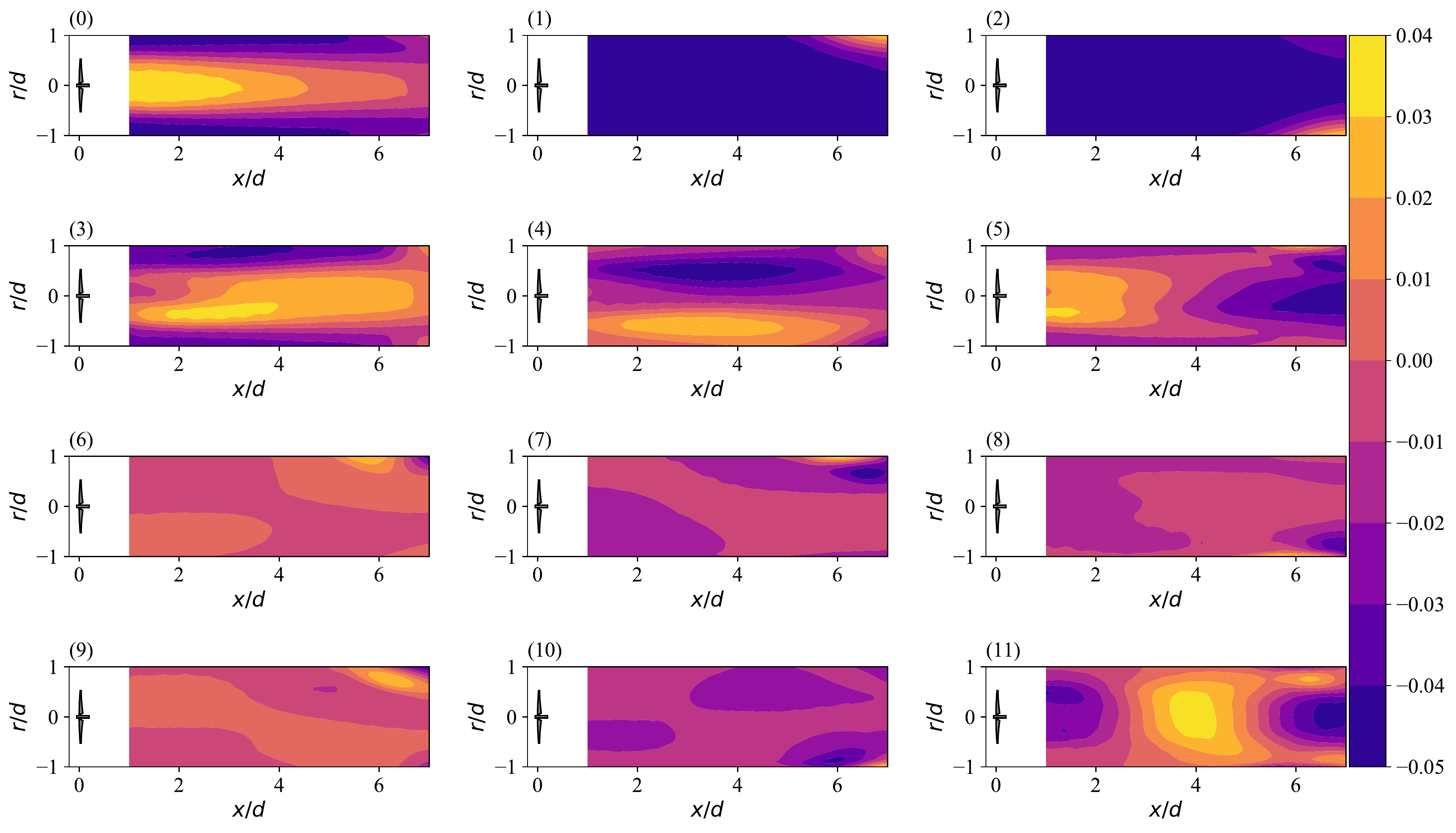}
    \caption{First 12 most energetic POD modes obtained from the LiDAR dataset.}
    \label{fig:PODmodes}
\end{figure}

The first 12 most energetic POD modes are reported in Figure \ref{fig:PODmodes}. POD mode 0 resembles the ensemble average of the interpolated wake velocity fields, which is shown in Figure \ref{fig:Mean}(a). Therefore, this POD mode has an evident physical contribution to the wake morphology. In contrast, POD modes 1 and 2 represent the corrections performed through the interpolation of the LiDAR data with the Matlab function inpaint-nans \citep{DErrico2004}, and, thus, they are ignored being a numerical artifact rather than a physical feature for the POD basis. POD modes 3 and 4 seem to indicate a modulation in the transverse direction over the shape provided through the ensemble mean (POD mode 0). In other words, POD modes 3 and 4 can represent contractions or expansions of the wake in the transverse direction, and non-symmetric wake conditions that were already observed in previous works for the near wake under stable atmospheric conditions \cite{ZhanWE2020,ZhanWES2020}. Similarly, POD modes 5 and 11 seem to indicate contractions or extensions of the wake in the streamwise direction. The remaining POD modes shown in Figure \ref{fig:PODmodes} seem to indicate similar wake distortions, yet with slightly larger wavelengths, which can be considered as sub-harmonics of the above-mentioned wake modulations. Based on this qualitative analysis of the POD modes, which we understand is speculative rather than based on quantitative characteristics, the truncated POD basis selected for this work includes only POD modes 0, 3, 4, 5, and 11, which allows reconstructing 44.8\% of the overall energy of the velocity covariance matrix. 

\begin{figure}[h!]
    \centering
    \includegraphics[width=\textwidth]{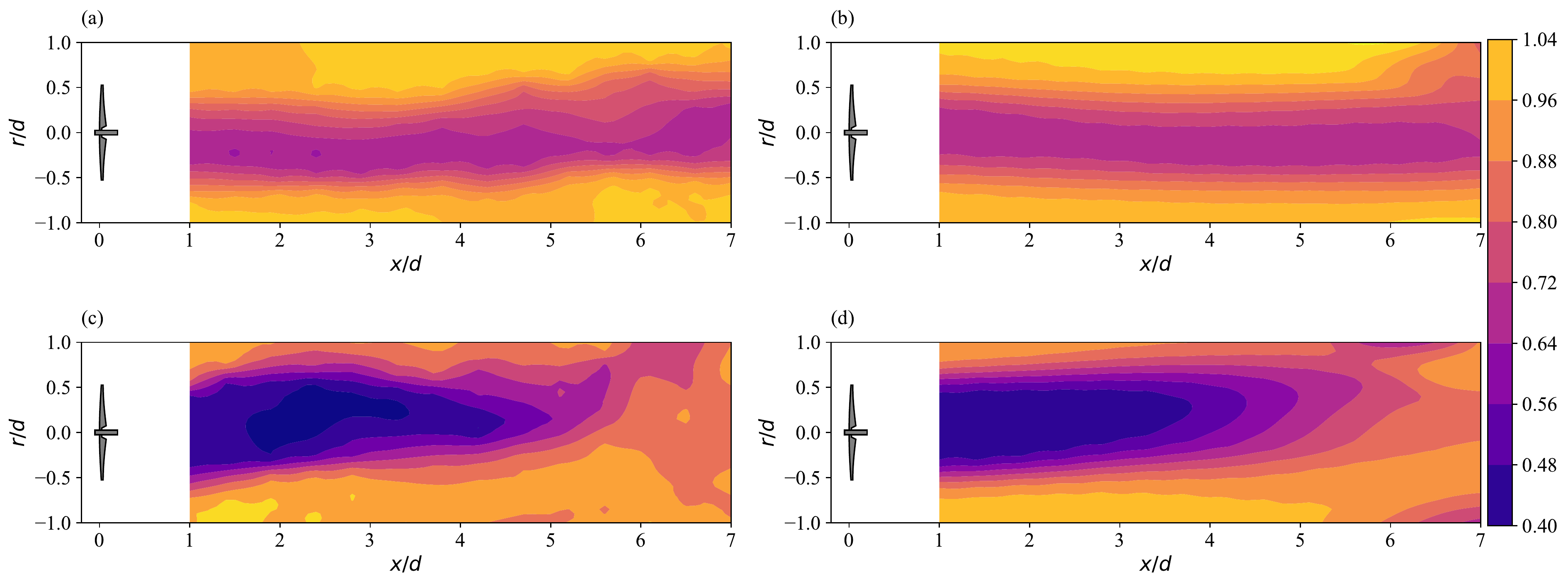}
    \caption{Reconstruction of the non-dimensional velocity field, $U/U_\infty$, through the selected truncated POD basis. The top row shows a case for $U_{hub}$=11.8 m/s, $TI$=5.1\%, $P$=2,306 kW, $Ri_B$=0.003; where LiDAR data is reported in (a) and POD-reconstructed velocity field in (b). The bottom row shows a case for $U_{hub}$=8.35 m/s, $TI$=14.3\%, $P$=1,279 kW, $Ri_B$=-0.002; where LiDAR data is reported in (c) and POD-reconstructed velocity field in (d).}
    \label{fig:PODreconst}
\end{figure}

Examples of the approximation performed on the LiDAR dataset through the truncated POD basis selected for this work are reported in Figure \ref{fig:PODreconst}. Two LiDAR snapshots collected during operations in Region 3 of the power curve and stable atmospheric condition, and Region 2 of the power curve and convective atmospheric condition are reported in Figures \ref{fig:PODreconst}(a) and \ref{fig:PODreconst}(c), respectively. Experimental details of those measurements are reported in the caption of that figure. The corresponding POD-approximated wake velocity fields are reported in Figures \ref{fig:PODreconst}(b) and \ref{fig:PODreconst}(d), respectively. For both cases, a good approximation of the wake morphology is achieved, especially in the near-wake region, while noticeable differences are observed at the most downstream locations. In general, the wake morphology is smoothed throughout the spatial domain. This seems an affordable drawback of the POD data compression, considering the reduced computational costs of the following cluster analysis achieved by reducing the dimensionality of each snapshot from $pq=2,501$ down to 5.   

\section{k-means clustering of the wind LiDAR data}
\label{sec:KMeans}
Clustering in general refers to a very broad set of techniques for finding subgroups from a dataset, which are referred to as clusters. The aim of clustering a dataset is to partition it into distinct sub-groups where samples sharing similar features belong to the same cluster and are segregated from those characterized by different features \cite{jain1999data}.

For this work, the $k$-means algorithm is used, which is a simple approach for partitioning a dataset into $K$ distinct, non-overlapping clusters \cite{Likas2003}. To perform $k$-means clustering, the desired number of clusters, $K$, must be provided as input. The standard method used to perform $k$-means clustering is an iterative algorithm. Firstly, a random integer (from 1 to $K$) is assigned to each sample as an initial cluster assignment. Next, for each of the $K$ clusters, the centroid of the cluster is computed and each observation is assigned to the cluster that it is closest to. These two steps are repeated until convergence in the cluster association is achieved \cite{James_2013B}. 

After the generation of clusters, a silhouette analysis is performed for each cluster to reject samples considered as outliers for their respective cluster \cite{Rousseeuw1987}. The silhouette analysis is performed by quantifying for each sample its distance from the respective cluster centroid through the $L_2$-norm, and the distances from the centroids of the remaining clusters. The silhouette coefficient has a range between -1 and 1, where 1 corresponds to the respective cluster centroid, 0 to the boundary of the cluster associated with the sample, and a negative number indicates that the sample might be an outlier for the considered cluster. In this work, all the samples with a non-positive silhouette coefficient are rejected for further analyses.

For our study, the only inputs provided for the cluster analysis are the time-series of the coefficients associated with the five POD modes selected to approximate the LiDAR dataset. It is noteworthy that no input is related neither to the wind turbine operative conditions nor to the wind/atmospheric conditions, as for previous cluster analyses of wind turbine wake measurements \citep{ZhanWE2020,ZhanWES2020}. The $k$-means outputs are the cluster centers, namely the representative realization for each cluster, and labeling for each LiDAR snapshot to its respective cluster. 

One of the challenges with using the $k$-means clustering algorithm is the choice of the number of clusters, $K$. This decision process is facilitated through the evaluation of the ``inertia curve'', which shows the relevance of each cluster within the dataset. It should be considered that increasing the number of clusters, $K$, might lead to the partitioning of clusters with a larger data population, rather than identifying different data features. Therefore, for this work, we have applied a hierarchical clustering approach, where the data are classified according to a cluster tree, denoted as dendogram \cite{Forina2002}. The dataset and the generated clusters (denoted as nodes) are partitioned into more successor sub-groups. Finally, all the nodes and sub-groups are nested and organized with a tree-like structure to provide a more physical and meaningful classification of the data. For the sake of efficiency, the number of the groups generated from a single node should be limited to clusters representing distinct physical features, i.e. wake characteristics in terms of velocity deficit and recovery rate.

\begin{figure}[b!]
    \centering
    \includegraphics[width=\textwidth]{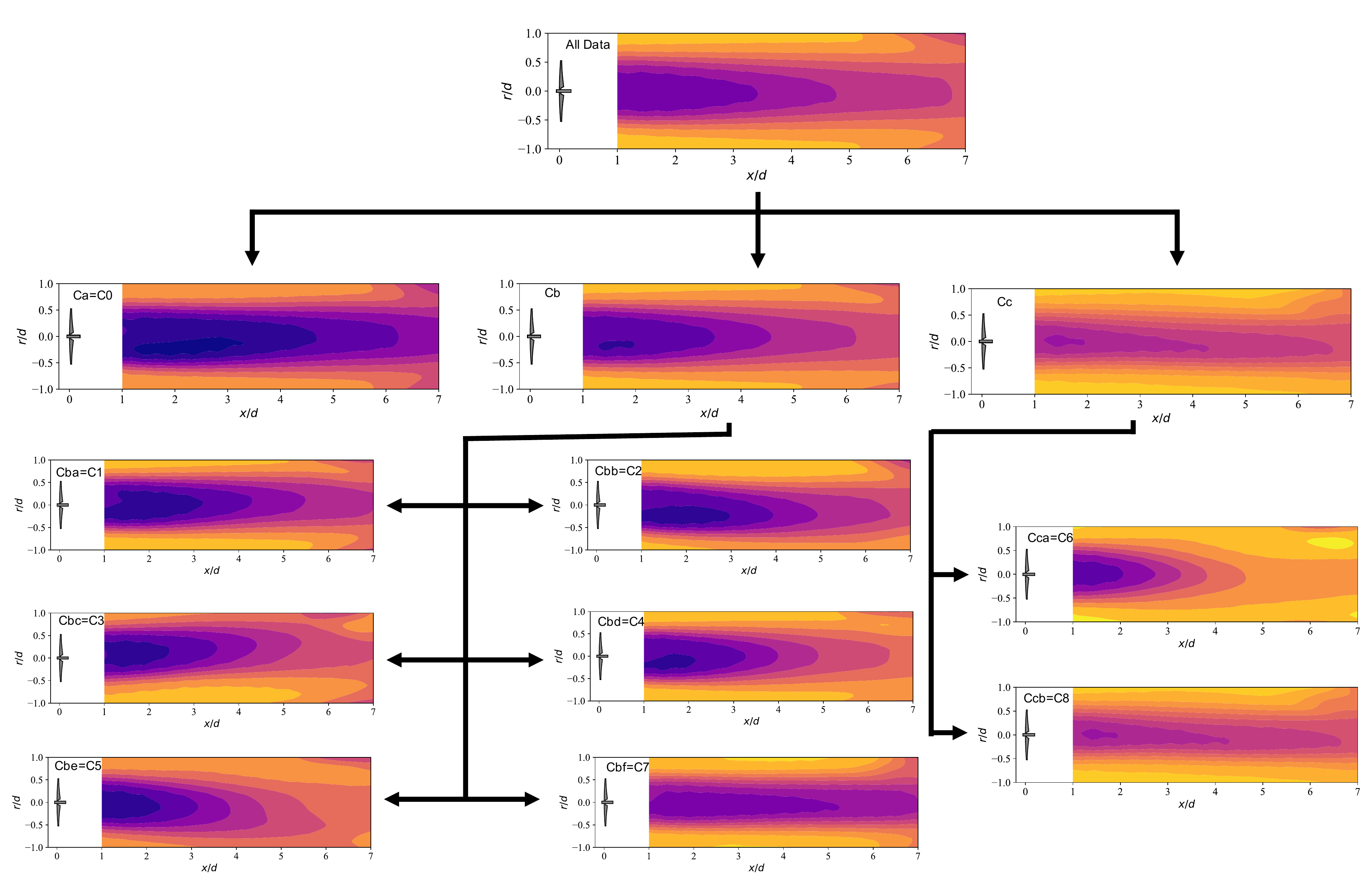}
    \caption{Dendogram of the LiDAR dataset reporting the cluster centroids. The colorbar (not reported) is the same as for Figure \ref{fig:PODreconst}.}
    \label{fig:ClusterCenters}
\end{figure}

After a preliminary cluster analysis, we decided to perform a first partitioning of the LiDAR dataset into three clusters, denoted as $Ca$, $Cb$, and $Cc$. The centroids of these clusters are reported in Figure \ref{fig:ClusterCenters}, which are obtained by adding the centroids of each principal component multiplied by their respective POD mode. In other words, these cluster centroids are a proxy for the ensemble average of the LiDAR velocity fields belonging to the same cluster. 

With this initial clustering, the dataset has been partitioned within groups with similar populations, as reported in the column ``Occurrence'' of Table \ref{tab:Clusters}. Figure \ref{fig:ClusterCenters} shows that the main differences among the centroids of the three clusters are related to the wake velocity deficit and the downstream extent of the wakes. 
\begin{table}[b!]
    \centering
    \begin{tabular}{||c|c|c|c|c|c|c|c|c|c|c|c||}
    \hline
    $\boldsymbol{Cluster~Label}$ & $\boldsymbol{C0}$ & - & - & $\boldsymbol{C1}$ & $\boldsymbol{C2}$ & $\boldsymbol{C3}$ & $\boldsymbol{C4}$ & $\boldsymbol{C5}$ & $\boldsymbol{C6}$ & $\boldsymbol{C7}$ & $\boldsymbol{C8}$\\
    \hline
    \hline
    $\boldsymbol{Initial~Label}$ & $Ca$ & $Cb$ & $Cc$ & $Cba$ & $Cbb$ & $Cbc$ & $Cbd$ & $Cbe$ & $Cca$ & $Cbf$ & $Ccb$\\
    \hline
    $\boldsymbol{Occurrence}$ &  28.7\%  & 40.4\% & 30.8\% & 9.1\% & 5.3\%  & 4.8\%  & 6.2\%  & 3.3\%  & 4.0\%  & 8.2\%  & 23.4\%\\ 
    \hline
    $\boldsymbol{Inertia}$ & 35.2 & 19.1 & 22.3 & 8.3 & 10.2 & 17.1 & 9.1 & 7.7 & 15.6 & 12.8 & 19.9\\ 
    \hline
    $\boldsymbol{x_{tr}/d}$ & 2.9 & - & - & 2.2 & 2.3 & 1.5 & 1.9 & 1.5 & 1.4 & 2.3 & 1.9\\
    \hline
    $\boldsymbol{A_u}$ & 1.00 & - & - & 1.38 & 1.32 & 1.55 & 2.17 & 2.20 & 3.00 & 0.59 & 0.49\\
    \hline
    $\boldsymbol{N_u}$ & 0.46  & -  &  - & 0.78  & 0.83 & 1.10 & 1.29 & 1.45 & 2.26 & 0.23 & 0.34\\ 
    \hline
    $\boldsymbol{\nu_t/U_\infty/d}$ & 0.0034 & - & - & 0.0073 & 0.0105 & 0.0223 & 0.0152 & 0.0305 & 0.1000 & 0.0010 & 0.0021\\
    \hline
    $\boldsymbol{C_P}$ & 0.35 & - & - & 0.40 & 0.40 & 0.39 & 0.41 & 0.46 & 0.43 & 0.30 & 0.26\\
    \hline
    $\boldsymbol{C_T^{SCADA}}$& 0.40 &  - & - & 0.46 & 0.46 & 0.45 & 0.47 & 0.55 & 0.51 & 0.33 & 0.28\\
    \hline
     $\boldsymbol{C_T^{AD}}$ & 0.95 & - & - &    0.90 &    0.89 &    1.01 &    0.87  &  1.14 & 0.81 &    0.70  &  0.57\\
     \hline
     $\boldsymbol{k^{*}}$ & 0.017  & - & - & 0.029 & 0.017 & 0.050 & 0.023 & 0.071 & 0.050 & 0.012 & -0.004\\
     \hline
     $\boldsymbol{\sigma_{U_\infty}/U_\infty}$ & 0.04  & - & - & 0.066 & 0.063 & 0.085 & 0.078 & 0.082 & 0.088 & 0.039 & 0.053\\
     \hline
    \end{tabular}
    \caption{Cluster parameters.}
\label{tab:Clusters}
\end{table}

These three clusters ($Ca$, $Cb$, $Cc$) are now considered as nodes of the dendogram, and, thus, each cluster is re-processed through the $k$-means algorithm to generate further sub-groups of the LiDAR dataset. The cluster analysis of $Ca$ has led to the generation of subgroups (not shown here for the sake of brevity) with very similar flow characteristics. Therefore, cluster $Ca$ is considered a definitive cluster for our study, which is ultimately labeled as $C0$. In contrast, the sub-cluster analysis of node $Cb$ leads to the identification of subgroups with a gradually varying downstream extent of the wake. In  Figure \ref{fig:ClusterCenters}, these sub-groups are indicated from $Cba$ to $Cbf$. A total number of six sub-groups is selected after a preliminary analysis to cover the variability of the wake velocity field through the cluster centers while limiting the number of clusters generated. It is noteworthy that the centroid of cluster $Cbf$ is characterized by a significantly lower velocity deficit and a longer downstream extent, which makes it more similar to the cluster centroid of node $Cc$ in terms of wake features.

For the cluster analysis of the node $Cc$, two clusters are deemed sufficient to identify the main wake topologies encompassed within this sub-dataset. Specifically, the sub-group $Cca$ is characterized by a strong near-wake velocity deficit and fast recovery, while the sub-group $Ccb$ preserves wake features similar to those of the respective node $Cc$.

\begin{figure}[hp]
\centering
\includegraphics[width=.66\textwidth]{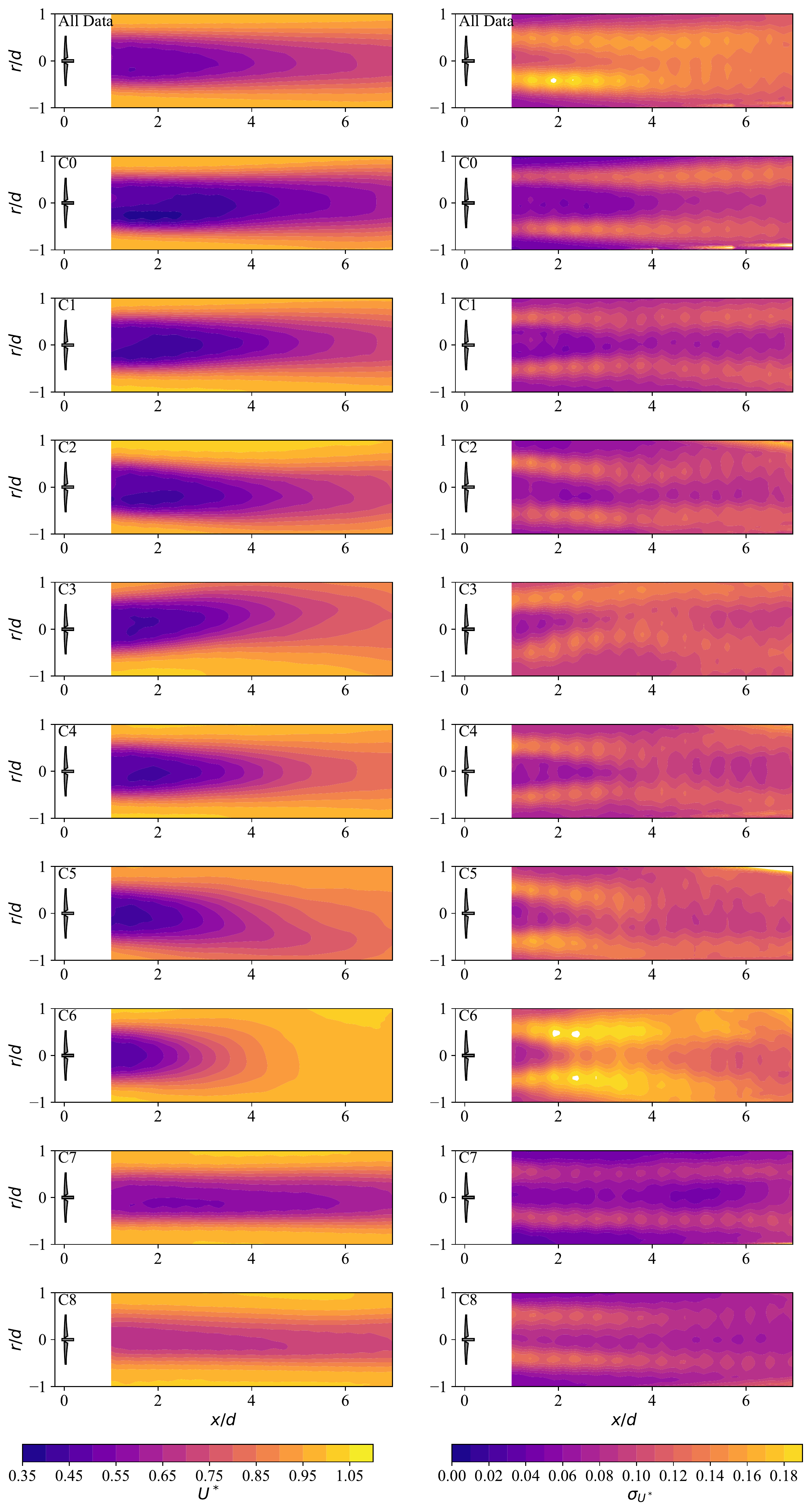}
\caption{Ensemble statistics of the LiDAR wake measurements for the various clusters: (left column) mean non-dimensional velocity fields, $U/U_\infty$; (right column) standard deviation  of the non-dimensional velocity fields, $\sigma_U/U_\infty$.}
\label{fig:ClusterStats}
\end{figure}

In summary, through the $k$-means algorithm and the dendogram approach, nine clusters are generated from the initial LiDAR dataset, which are numbered from $C0$ up to $C8$. The ensemble statistics in terms of average and standard deviation for the entire dataset and the various clusters are reported in Figure \ref{fig:ClusterStats}. As mentioned above, the ensemble mean fields resemble the cluster centroids already reported in Fig. \ref{fig:ClusterCenters}, yet avoiding data distortion introduced by the interpolation of the LiDAR data required for the application of the POD. The ensemble statistics highlight the broad variability of the wake velocity field for different settings of the turbines and meteorological conditions, both for the mean and standard deviation. Furthermore, it is noteworthy that clusters $C2$ and $C3$ show evident skewing of the wake from the $x$-axis representing the wake direction. This particular wake morphology, which can be ascribed to a certain yaw misalignment of the rotor disc from the mean wind direction \cite{Bastankhah2016}, will be analyzed more in detail in the following section. 

\section{Analysis of the clustered datasets}
\label{sec:Analysis}
\subsection{Mean wake velocity field for the various clusters}
\label{subsec:Mean}
\begin{figure}[t!]
    \centering
    \includegraphics[width=.6\textwidth]{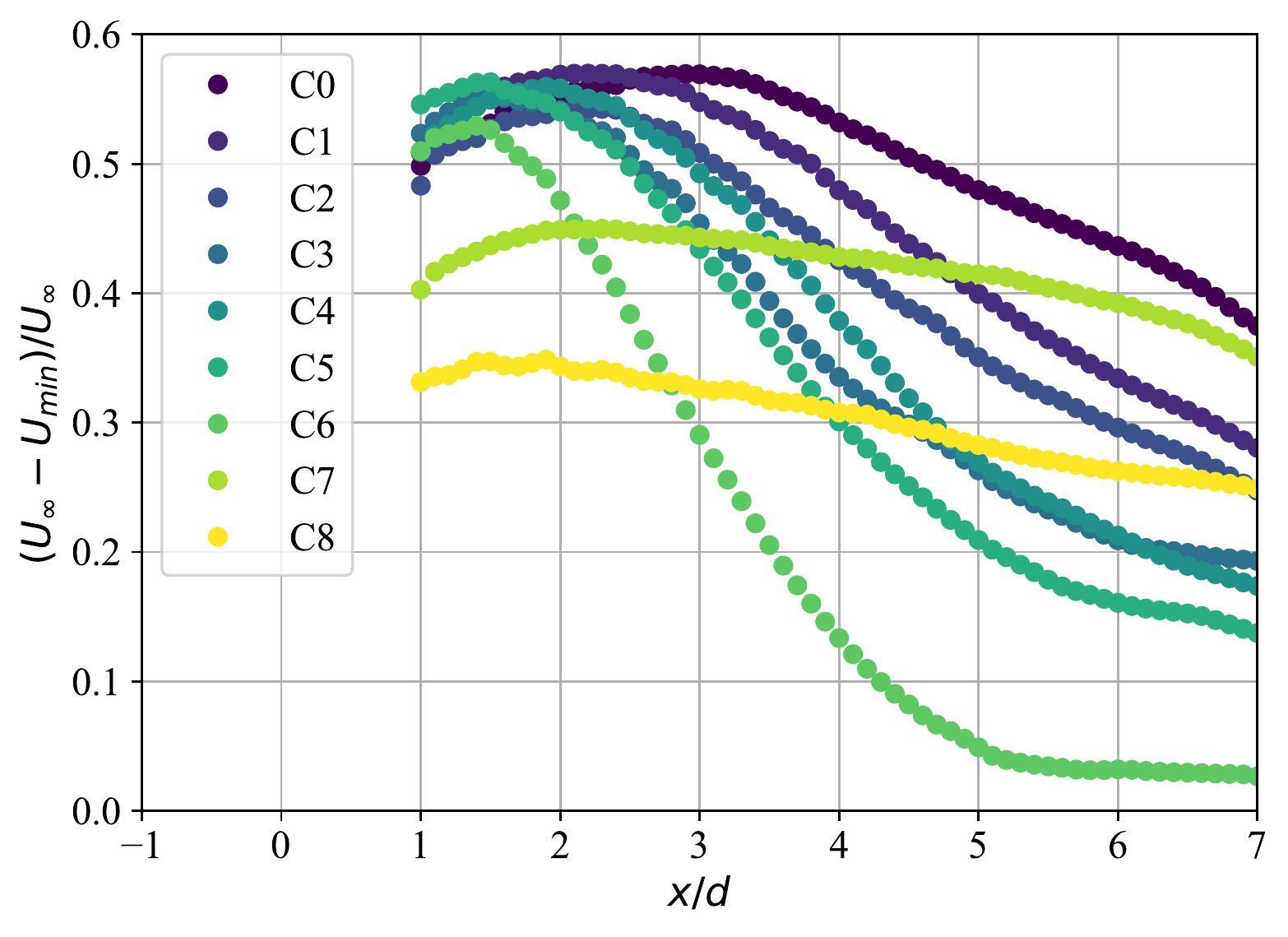}
    \caption{Maximum velocity deficit, $\Delta U_{min}$, as a function of the streamwise coordinate for the various clusters.}
    \label{fig:ClusterUmin}
\end{figure}

For characterizing the mean wake velocity obtained for the various clusters, the maximum velocity deficit, $\Delta U_{min}/U_\infty=(U_\infty-U_{min})/U_\infty$, ($U_\infty$ is the hub-height freestream wind speed and $U_{min}$ is the minimum streamwise velocity at a given downstream distance), as a function of the downstream position is reported in Figure \ref{fig:ClusterUmin}. This plot emphasizes even more clearly the variability in velocity deficit and wake persistence along the downstream direction that can be observed during typical operations of a wind turbine. For the clusters from $C0$ up to $C6$, the velocity deficit in the very near wake is comparable, which suggests that the thrust coefficient of the turbine might be very similar for these clusters. Ranging from $C0$ to $C6$, we observe generally faster recovery of the velocity deficit. The velocity deficit at $x/d=1$ is lower for cluster $C7$, and even more for $C8$. This wake feature may indicate that clusters $C7$ and $C8$ belong to turbine operations with an active pitch control of the turbine blades and, thus, reduced rotor thrust coefficient. For $C7$ and $C8$, the recovery of the velocity deficit is significantly slower than for the remaining clusters.  

Starting from the most upstream location, i.e. $x/d=1$, the velocity deficit generally increases in the downstream direction, which indicates that the turbine forcing, mainly through the generation of a streamwise pressure gradient, is still acting on the incoming flow. The downstream location where the maximum velocity deficit occurs, $x_{tr}$, can be associated with the transition between the near wake and the far wake \cite{Vermeulen1980}. This analysis shows that the near-to-far wake transition generally moves upstream switching from cluster $C0$ to $C6$, while it seems roughly invariant for $C7$ and $C8$ (Table \ref{tab:Clusters}). 

To provide a more quantitative analysis of the mean wake velocity fields associated with the various clusters, the maximum velocity deficit, $\Delta U_{min}/U_\infty$, is fitted through the following power law for $x/d\geq3$ \citep{IungoJTECH2014,ZhanWE2020}:
\begin{equation}
    \label{eq:Umin}
    \frac{\Delta U_{min}}{U_\infty}=1-A_u \left(\frac{x}{d} \right)^{-N_u},
\end{equation}
where $A_u$ and $N_u$ represent the velocity deficit at $x/d=1$ and the wake-recovery rate, respectively. As reported in Table \ref{tab:Clusters}, the parameter $A_u$ generally increases spanning from the cluster $C0$ up to $C6$ between values 1 and 3, respectively. The fitted values of $A_u$ are significantly larger than the actual maximum velocity deficit at $x/d=1$ because of a different variability of the mean velocity field in the near wake, i.e. for $x\approx<x_{tr}$ (see Figure \ref{fig:ClusterUmin}). Similarly, the parameter $N_u$ increases from 0.46 for cluster $C0$ up to 2.26 for $C6$. Therefore, the variability of $N_u$ corroborates a faster wake recovery spanning from $C0$ up to $C6$.

Completely different wake features are quantified for clusters $C7$ and $C8$, for which the reduced near-wake velocity deficit leads to $A_u$ equal to 0.59 and 0.49, respectively, and very low wake recovery rates than for the previous clusters, i.e. 0.23 and 0.34, respectively.

As mentioned above, the velocity deficit in the near-wake is strictly connected with the rotor thrust coefficient, $C_T$ \cite{Iungo2018,ZhanWES2020}. Estimates of $C_T$ from the mean wake velocity field can be generated by considering a 1D axisymmetric stream tube, and coupling mass conservation and the momentum equation in the streamwise direction \cite{Batchelor1967,Pope2000}:
\begin{equation}
\label{eq:CtAD}
    C_T^{AD} =16   \int_0^\infty \frac{U}{U_\infty} \left(1-\frac{U}{U_\infty}\right)\frac{r}{d}~d\frac{r}{d} 
\end{equation}
Eq. \ref{eq:CtAD} is solved numerically for the mean velocity fields $U/U_\infty$ of each cluster at the various downstream locations. The estimates of the thrust coefficient through the actuator disk theory, $C_T^{AD}$, are obtained for each cluster at the location of the maximum velocity deficit, $x_{tr}$ (see Table \ref{tab:Clusters}). These results confirm that the thrust coefficient for clusters $C0$-$C6$ is substantially larger (between 0.81 and 1.14) than for $C7$ and $C8$ (0.7 and 0.57, respectively).

It is noteworthy that for the clusters $C0$-$C6$, the variability of $C_T^{AD}$ does not reflect the variability observed for the parameter $A_u$. Indeed, the quantification of $A_u$ is affected not only by the rotor thrust in the near wake but also by the location of the transition from the near and far wake, $x_{tr}$. Therefore, we consider  $C_T^{AD}$ a more reliable parameter for the characterization of the rotor thrust and the near-wake velocity field.  

The recovery rate of the ensemble mean velocity field for each cluster is further characterized by calculating the optimal turbulent eddy-viscosity, $\nu_T$, for an axisymmetric Reynolds-averaged Navier Stokes (RANS) model \cite{Iungo2018} (the RANS code is publicly available \cite{GRANS}). Specifically, for each cluster, the mean velocity measured at $x/d=1$ is provided as the inlet condition for the RANS model, while a constant $\nu_T$ is optimized by minimizing the $L_2$-norm of the difference between the LiDAR mean velocity field and the velocity field predicted through the RANS model. For the optimization, the sequential quadratic programming implemented in Matlab through the fmincon function is adopted \cite{Fletcher1987} with termination tolerance on the first optimality of $10^{-6}$. The estimates for $\nu_T$ show similar trends as for the parameter $N_u$, namely the lowest non-dimensional values are obtained for the clusters $C7$ and $C8$ (0.0010 and 0.0021, respectively), reflecting the very slow wake recovery and long downstream extent of the respective mean velocity fields. A larger $\nu_T/U_\infty/d$ is estimated for $C0$ (0.0034), which is characterized by a significant downstream extent, yet larger near-wake velocity deficit. For the remaining clusters, $\nu_T/U_\infty/d$ increases up to 0.1, reflecting a gradually shorter downstream extent and faster wake recovery.   

The majority of the clusters are characterized by a quasi-symmetric wake. However, the clusters $C2$ and $C3$ show a significant transverse deflection of the wake center, reaching maximum values, $\delta_\infty$, of about -0.19$d$ and 0.33$d$, respectively (Figure \ref{fig:ClusterStats}). This wake feature can be ascribed to a misalignment of the rotor axis with the mean wind direction of a certain angle $\gamma$, similarly to a wind turbine operating under yawed conditions \cite{Howland2016,Gebraad2016,Bastankhah2016}. For a given asymptotic wake deflection, the rotor yaw angle, $\gamma$, can be estimated as \cite{Bastankhah2016}:
\begin{equation}
\begin{split}
 \frac{\delta_\infty}{d}=\frac{0.3~\gamma}{\cos{\gamma}} \left(1-\sqrt{1-C_T \cos{\gamma}}\right) \left[\frac{\cos{\gamma}(1+\sqrt{1-C_T})}{\sqrt{2}(\alpha^*TI+\beta^*(1-\sqrt{1-C_T}))}+\right.\\
\left.+\frac{1}{14.7}\sqrt{\frac{\cos{\gamma}}{k^{*2} C_T}}(2.9+1.3 \sqrt{1-C_T}-C_T) \times \ln{\left( \frac{1.6+\sqrt{C_T}}{1.6-\sqrt{C_T}}\right)} \right],
\end{split}
\end{equation}
where $\alpha^*=2.32$, $\beta^*=0.154$, and $k^*$ is the linear growth rate of the wake width. Using as thrust coefficient the values obtained through the symmetric actuator disk theory, $C_T^{AD}$, and $k^*$ estimated through the Gaussian model \cite{Bastankhah2014} (both parameters are reported in Table \ref{tab:Clusters}), these asymptotic wake deflections would correspond to a rotor yaw angle, $\gamma$, of $-3.2^\circ$ and $5.6^\circ$ for $C2$ and $C3$, respectively.

This result is interesting because the cluster analysis of the LiDAR data has enabled the identification of these operations with a systematic yaw misalignment, leading to a noticeable deflection of the wind turbine wakes, which might be important for the power efficiency of the entire wind farm. These operations under yaw misalignment have a total occurrence of about 10\% of the duration of the LiDAR observations, which is not a negligible number. The detection of these off-design conditions would not have been possible by generating bins of the LiDAR dataset by imposing thresholds on the various wind and turbine parameters, as done for previous works \cite{ZhanWE2020,ZhanWE2020}; in contrast, this result has been achieved only thanks to the data-driven approach of the k-means clustering algorithm.  

\subsection{Effects of the clustering analysis on the meteorological and SCADA data}
\label{subsec:MetSCADA}
\begin{figure}[t!]
    \centering
    \includegraphics[width=\textwidth]{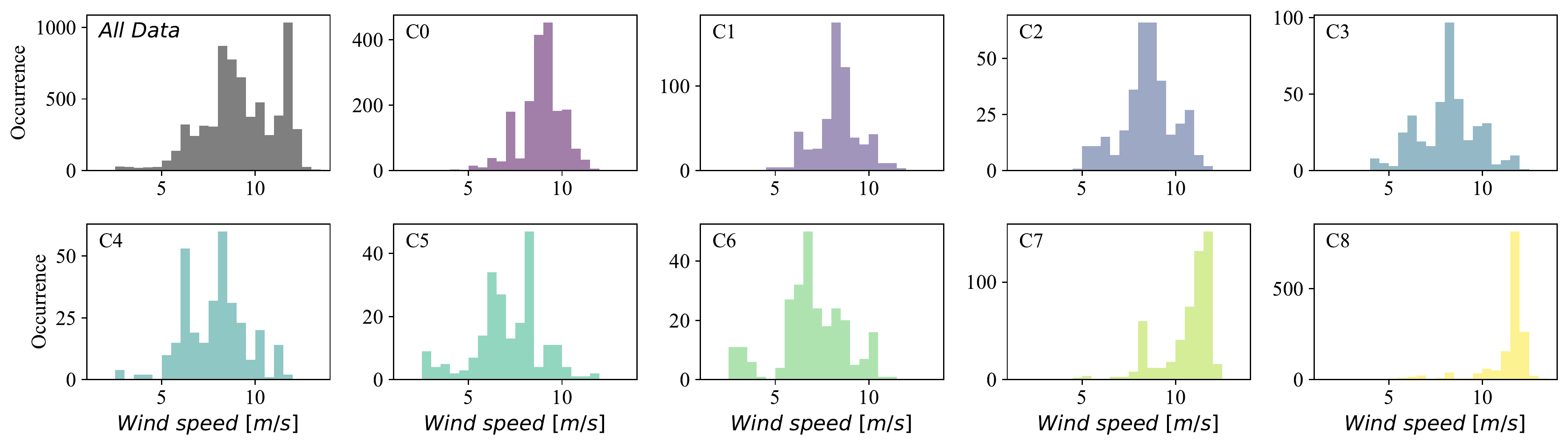}
    \caption{Histograms of the hub-height wind speed recorded through the SCADA for the various clusters.}
    \label{fig:ClusterHistWS}
\end{figure}

For the various clusters, we now analyze the statistics of SCADA and meteorological parameters, which were collected simultaneously to the LiDAR data. It is noteworthy that these parameters were not fed into the cluster analysis and, thus, their variability has not affected the generation of the data clusters. The histogram of the mean hub-height wind speed recorded through the SCADA is reported in Figure \ref{fig:ClusterHistWS} for the entire dataset and the various clusters, while the 25-$th$, 50-$th$, and 75-$th$ percentiles for the entire dataset and the various clusters are reported in Table \ref{tab:SCADAmetClusters}. Considering that the rated wind speed of the turbines under investigation is 11.5 m/s, then it becomes evident that the clusters from $C0$ up to $C6$ belong to operations in region 2 of the power curve, i.e. for hub-height wind speed between cut-in and rated wind speeds, while the cluster $C8$ belongs to operations in region 3, i.e. between rated wind speed and cut off wind speed, and the cluster $C7$ seems to encompass operations at the transition between region 2 and region 3, typically denoted as region 2.5 \cite{Burton2011}.

\begin{table}[h!]
    \centering
    \begin{tabular}{||c|c|c|c|c|c|c|c|c|c|c||}
    \hline
   $\boldsymbol{Parameter}$ & $\boldsymbol{All~Dataset}$ & $\boldsymbol{C0}$ & $\boldsymbol{C1}$ & $\boldsymbol{C2}$ & $\boldsymbol{C3}$ & $\boldsymbol{C4}$ & $\boldsymbol{C5}$ & $\boldsymbol{C6}$ & $\boldsymbol{C7}$ & $\boldsymbol{C8}$\\
\hline
\hline
$\boldsymbol{U_{hub}^{p25} [m/s]}$ & 8.1 & 8.4 & 7.9 & 7.9 & 6.4 & 7.3 &6.1 & 6.1 & 10.2 & 11.3\\
\hline
$\boldsymbol{U_{hub}^{p50} [m/s]}$ & 9.1 & 9.0 & 8.3 & 8.5 & 8.1 & 8.2 &7.0 & 6.9 & 11.0 & 11.7\\
\hline
$\boldsymbol{U_{hub}^{p75} [m/s]}$ & 11.0 & 9.6 & 8.9 & 9.1 & 8.8 & 9.1 &8.2 & 8.2 & 11.6 & 11.9\\
\hline
\hline
$\boldsymbol{Power^{p25} [kW]}$ & 1173 & 1233 & 1094 & 1098 & 664 & 874 &591 & 600 & 1990 & 2265\\
\hline
$\boldsymbol{Power^{p50} [kW]}$ & 1530 & 1440 & 1279 & 1377 & 1173 & 1261 &887 & 798 & 2239 & 2304\\
\hline
$\boldsymbol{Power^{p75} [kW]}$ & 2235 & 1680 & 1494 & 1690 & 1548 & 1652 &1270 & 1216 & 2293 & 2309\\
\hline
\hline
$\boldsymbol{Pitch^{p25} [deg]}$ & -1.93 & -1.99 & -1.97 & -1.97 & -1.93 & -1.96 &-1.93 & -1.93 & -0.24 & 0.68\\
\hline
$\boldsymbol{Pitch^{p50} [deg]}$ & -1.70 & -1.93 & -1.91 & -1.88 & -1.83 & -1.80 &-1.83 & -1.82 & 1.26 & 2.64\\
\hline
$\boldsymbol{Pitch^{p75} [deg]}$ & 0.89 & -1.50 & -1.76 & -1.49 & -1.59 & -1.48 & -1.56 & -1.57 & 1.99 & 4.38\\
\hline
\hline
$\boldsymbol{Ri_B^{p25}}$ & -0.0019 & 0.0008 & -0.0021 & -0.0022 & -0.0032 & -0.0030 & -0.0044 & -0.0044 & -0.0001 & -0.0004\\
\hline
$\boldsymbol{Ri_B^{p50}}$ & -0.0003 & 0.0016 & -0.0014 & -0.0012 & -0.0022 & -0.0021 & -0.0034 & -0.0034 & 0.0018 & 0.0003\\
\hline
$\boldsymbol{Ri_B^{p75}}$ & 0.0018 & 0.0026 & -0.0007 & 0.0008 & -0.0015 & -0.0017 & -0.0026 & -0.0025 & 0.0038 & 0.0012\\
\hline
\hline
$\boldsymbol{TI^{p25} [\%]}$ & 5.7 & 5.0 & 8.8 & 7.3 & 10.9 & 11.2 & 13.1 & 13.7 & 4.7 & 5.7\\
\hline
$\boldsymbol{TI^{p50} [\%]}$ & 8.6 & 5.8 & 10.7 & 10.5 & 12.7 & 12.9 & 15.1 & 15.9 & 5.2 & 7.6\\
\hline
$\boldsymbol{TI^{p75} [\%]}$ & 11.9 & 7.2 & 12.4 & 12.4 & 15.4 & 14.7 & 18.1 & 18.9 & 7.7 & 9.6\\
\hline
\hline
$\boldsymbol{\alpha^{p25}}$ & 0.09 & 0.16 & 0.05 & 0.07 & 0.04 & 0.04 & 0.04 & 0.05 & 0.21 & 0.20\\
\hline
$\boldsymbol{\alpha^{p50}}$ & 0.22 & 0.42 & 0.14 & 0.16 & 0.07 & 0.07 & 0.07 & 0.07 & 0.38 & 0.31\\
\hline
$\boldsymbol{\alpha^{p75}}$ & 0.39 & 0.52 & 0.33 & 0.35 & 0.13 & 0.16 & 0.12 & 0.12 & 0.45 & 0.37\\
\hline
\end{tabular}
\caption{Clustering of the SCADA and meteorological parameters. The superscript $p25$, $p50$, and $p75$ indicate the 25-$th$, 50-$th$, and 75-$th$ percentiles. The horizontal blocks of the table (from top to bottom) are the statistics of the hub-height wind speed, power capture, pitch angle, Bulk Richardson number, hub-height wind turbulence intensity, and wind shear exponent.}
\label{tab:SCADAmetClusters}
\end{table}

\begin{figure}[t!]
    \centering
    \includegraphics[width=\textwidth]{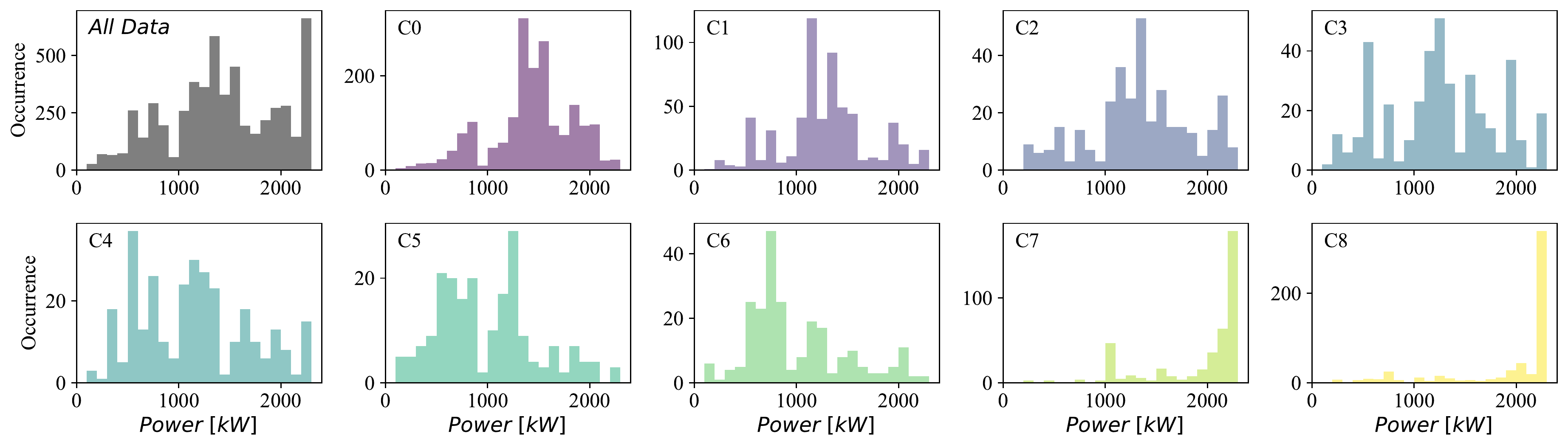}
    \caption{Histograms of the power capture recorded through the SCADA for the various clusters.}
    \label{fig:ClusterHistPower}
\end{figure}

\begin{figure}[t]
    \centering
    \includegraphics[width=\textwidth]{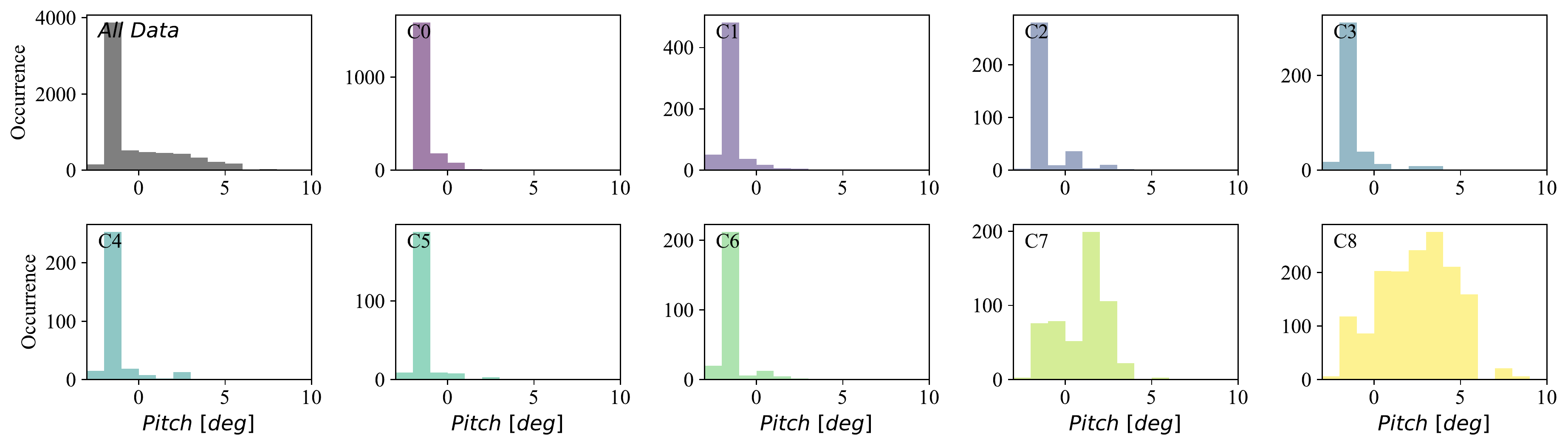}
    \caption{Histograms of the blade pitch angle recorded through the SCADA for the various clusters.}
    \label{fig:ClusterHistPitch}
\end{figure}

This classification of the various clusters based on the turbine operative conditions is corroborated from the respective histograms of the turbine power and blade pitch angle, Figures \ref{fig:ClusterHistPower} and \ref{fig:ClusterHistPitch}, respectively, and related statistics reported in Table \ref{tab:SCADAmetClusters}. Power capture for the clusters $C0$-$C6$ is lower than the rated power of 2,300 kW, while the blade pitch angle is generally between -2$^\circ$ and -1.5$^\circ$, which is the typical range for non-active pitch control of the turbine rotor. In contrast, power capture for the operations of clusters $C7$ and $C8$ belongs to operations with an active blade pitch control. Indeed, the statistics of the pitch angle for $C7$ and $C8$ are larger than for $C0$-$C6$ with median values of 1.26$^\circ$ and 2.64$^\circ$, which are associated with a median power capture of 2,239 kW and 2,304 kW, respectively. 

\begin{figure}[t]
    \centering
    \includegraphics[width=\textwidth]{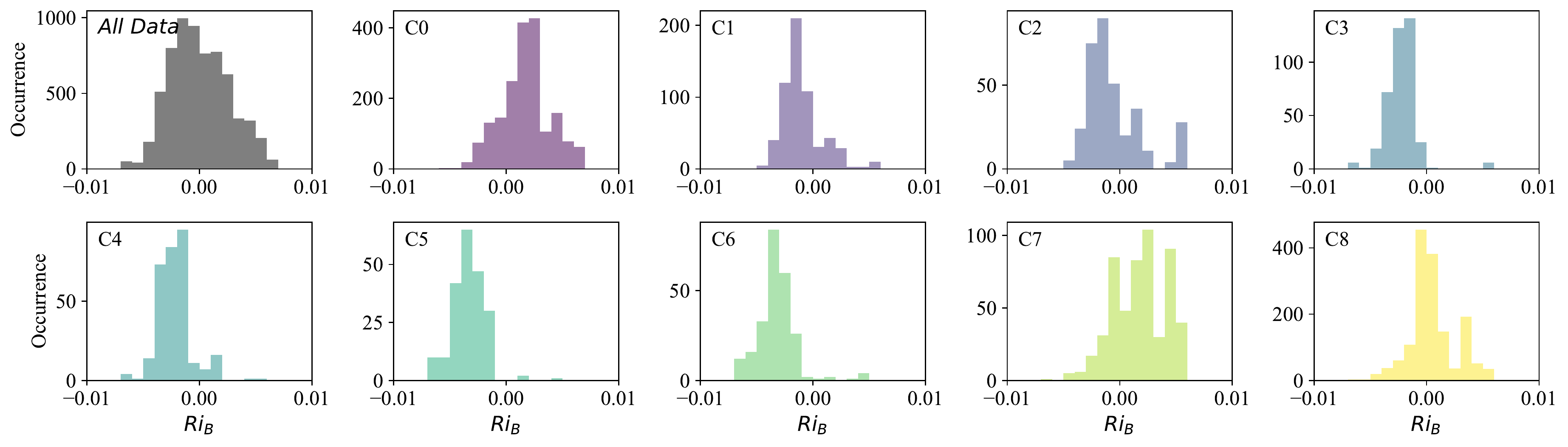}
    \caption{Histograms of the Bulk Richardson number calculated from the meteorological data for the various clusters.}
    \label{fig:ClusterHistRiB}
\end{figure}

\begin{figure}[t]
    \centering
    \includegraphics[width=\textwidth]{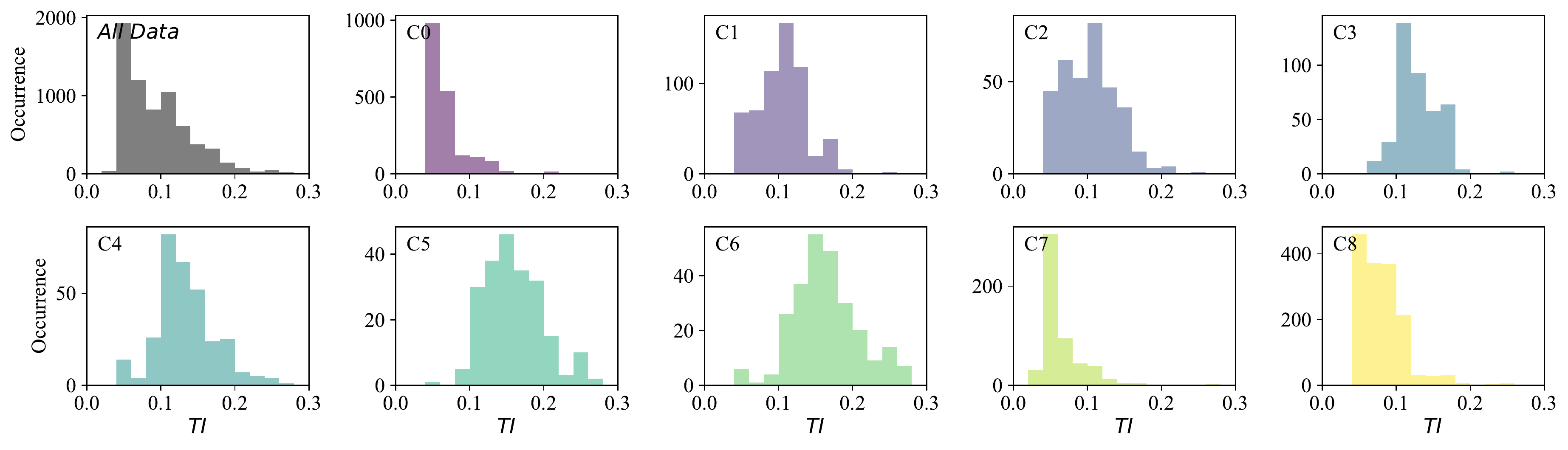}
    \caption{Histograms of the hub-height turbulence intensity recorded through the SCADA for the various clusters.}
    \label{fig:ClusterHistTI}
\end{figure}

\begin{figure}[h!]
    \centering
    \includegraphics[width=\textwidth]{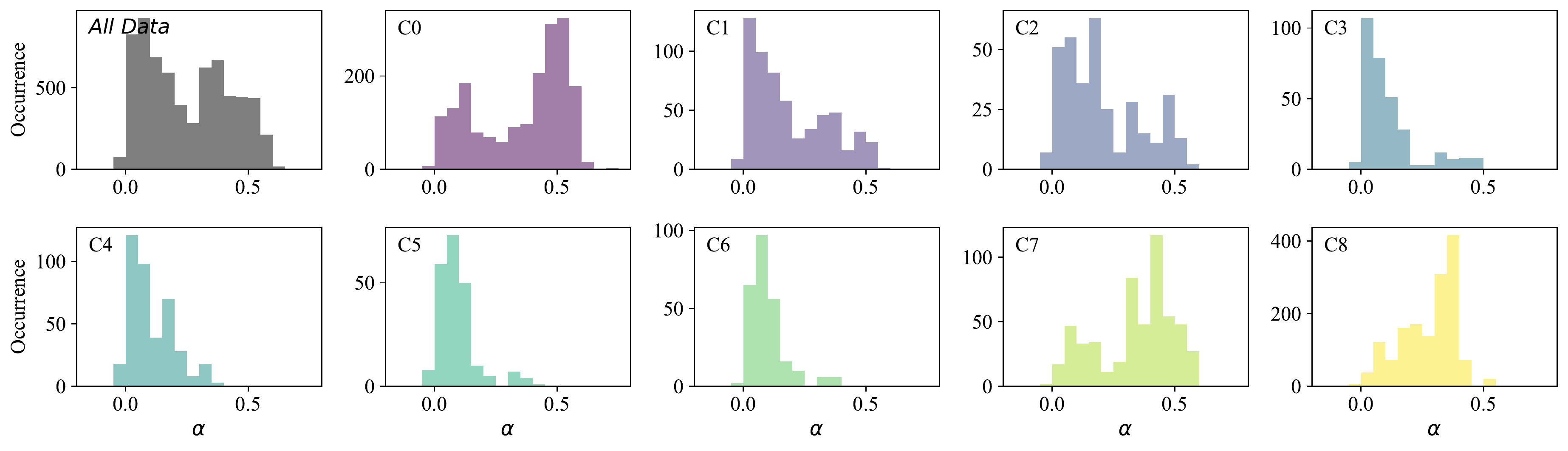}
    \caption{Histograms of the shear exponent of the incoming wind for the various clusters.}
    \label{fig:ClusterHistShearExp}
\end{figure}

Besides turbine settings, also atmospheric stability can significantly affect the characteristics and evolution of wind turbine wakes \cite{IungoJTECH2014}. To investigate the role of atmospheric stability on the variability observed through the ensemble statistics of the wake velocity field for the various clusters, histograms (Figures \ref{fig:ClusterHistRiB}, \ref{fig:ClusterHistTI}, \ref{fig:ClusterHistShearExp}) and percentiles (Table \ref{tab:SCADAmetClusters}) of the Bulk Richardson number, $Ri_B$, incoming wind turbulence intensity at hub height, $TI$, and shear exponent, $\alpha$, are also analyzed, respectively. Starting from the clusters owing operations in region 2 of the power curve ($C0$-$C6$), the Bulk Richardson number is generally positive for $C0$ (median $Ri_B$ of 0.0016), while it is typically negative for the remaining clusters $C1$-$C6$. Therefore, operations associated with $C0$ can occur under stable atmospheric conditions while $C1$-$C6$ under unstable conditions. Furthermore, the level of atmospheric instability increases ranging from $C0$ up to $C6$ (Table \ref{tab:SCADAmetClusters}).  

This analysis is confirmed from the statistics of the incoming wind turbulence intensity at hub height, $TI$, which has a median value of 5.8\% for $C0$ and the largest value of 15.9\% for $C6$. Similar conclusions are obtained from the statistics of the incoming-wind shear exponent, i.e. a larger value is reported for $C0$ (0.42), while the other clusters of region 2 can achieve a median value as low as 0.07.

The statistics of the meteorological parameters for the clusters owing to operations in the proximity of the rated wind speed, i.e. $C7$ and $C8$, indicate the prevalence of neutral-stable conditions. Indeed, the median value of the shear exponent is larger than 0.3, the median $TI$ is lower than 8\% and the statistics of the Bulk Richardson number indicate predominantly stable atmospheric conditions for $C7$ (median $Ri_B$=0.0018) and neutral atmospheric conditions for $C8$ (median $Ri_B$=0.0003).

To conclude this statistical survey of the SCADA and meteorological data partitioned through the cluster analysis, we calculate the median power coefficient, $C_P$, from the power capture of the turbine and the density-corrected hub-height wind speed (both parameters recorded by the SCADA). The minimum value of $C_P$ (0.26) is estimated for cluster $C8$, and it is slightly higher for $C7$ (0.3) (see Table \ref{tab:Clusters}). A reduction of $C_P$ with increasing wind speed in region 3 of the power curve is consistent with the operated blade pitching to keep power capture equal to the rater power of the turbine. The power coefficient, $C_P$ is larger for the clusters associated with operations in region 2 ($C0$-$C6$), with values noticeably larger for operations under unstable atmospheric conditions. This result corroborates a higher power efficiency of wind turbines under convective conditions, as already observed through previous wind farm studies \cite{ElAsha2017}.

By leveraging the actuator disk theory, it is possible to estimate the axial induction factor, $a$, from the power coefficient ($C_P=4a(1-a)^2$), and calculate, in turn, the thrust coefficient, $C_T^{SCADA}=4a(1-a)$. The latter is reported in Table \ref{tab:Clusters} for the various clusters, which confirms again an increasing axial induction of the turbine rotor moving from operations in region 3 to operations in region 2 under stable atmospheric conditions, and even further for operations in region 2 under unstable atmospheric conditions. More importantly, Table \ref{tab:Clusters} shows that estimates of the rotor thrust coefficient derived from the power data, $C_T^{SCADA}$, are about half of the respective values obtained from the mass conservation and streamwise momentum budget applied to the mean LiDAR wake measurements with axisymmetry assumption, $C_T^{AD}$. This feature was already highlighted in previous works \cite{Iungo2018}, confirming that accurate estimates of the rotor thrust coefficient for wake modeling and predictions of power capture are still elusive. 

Summarizing, this analysis of the SCADA and meteorological data for the various clusters, coupled with the analysis of the mean wake velocity field in the previous sub-section, indicates that the clusters $C0$-$C6$ belong to operations in region 2 of the power curve, $C7$ to operations at the interface between region 2 and region 3 with active pitch control, and $C8$ to operations in region 3. Regarding the wake variability connected with atmospheric stability, $C0$ and $C7$ capture operations under stable conditions for regions 2 and 3 of the power curve, respectively. The clusters $C1$-$C6$ belong to operations under unstable conditions with a gradually increasing level of instability, while LiDAR data of $C8$ were collected under quasi-neutral conditions.

\subsection{Wake-added turbulence intensity for the various clusters}
\label{subsec:WakeTI}
Wake turbulence intensity, $I_u=\sigma_u/U_\infty$, where $\sigma_u$ is the ensemble standard deviation of the streamwise velocity calculated through the cluster analysis (Section \ref{sec:KMeans}), affects directly the rotor aerodynamic forces ($\propto 1+I_u^2$), and turbine power capture ($\propto 1+3I_u^2+\gamma I_u^3$, where $\gamma$ is the skewness of the streamwise velocity) \cite{Binh2008,Chamorro2009}. Furthermore, turbulence intensity is very important in wind energy because it is typically used to model turbine fatigue loads \cite{Rosen1996}. Downstream to a turbine rotor, the wake-added standard deviation of the streamwise velocity, $\Delta u'$, is calculated from the standard deviation of the incoming wind velocity, $\sigma_{U_\infty}$, as \citep{Frandsen2007}:
\begin{equation}
    \Delta u'(x,r)=sign\left[\sigma^2_u(x,r)-\sigma^2_{U_\infty} \right]\sqrt{\left| \sigma^2_u(x,r)-\sigma^2_{U_\infty} \right|},
\end{equation}
where the operator $sign$ produces values equal to 1 or -1 if $\sigma^2_u$ is larger or smaller than $\sigma^2_{U_\infty}$, respectively. 

From hot-wire measurements carried out in a boundary-layer wind tunnel for the wake of a down-scaled wind turbine model, the wake-added turbulence intensity was observed to be either positive or negative over different wake areas and downstream locations \cite{Chamorro2009}. In the far wake, statistics of the wake-added turbulence intensity, such as mean, maximum, minimum, and average of only positive or negative added turbulence intensity, showed the decay of these parameters in the downstream direction according to power laws with exponent between -0.5 and -0.3. $\Delta u'$ reaches typically its maximum within the range $2 \lessapprox x/d \lessapprox 5$, which can coincide with the transition between the near- and far-wake regions. 

A simple model for predicting the maximum of the wake-added turbulence intensity, $\Delta u'_{max}/U_\infty$, was proposed by \citet{Crespo1996}:
\begin{equation}
    \frac{\Delta u'_{max}}{U_\infty}=\begin{cases}0.362 \left( 1-\sqrt{1-C_T }\right) & x/D \leq3\\
0.73 \left( \frac{1-\sqrt{1-C_T}}{2}\right)^{0.83} & x/D >3
\end{cases}
\label{eq:TImodel1}
\end{equation}
while the following model \citep{Frandsen2007} is used for the IEC61400-1 standard \citep{IEC}:
\begin{equation}
    \frac{\Delta u'_{max}}{U_\infty}=\frac{1}{1.5+0.8~C_T^{-0.5}x/d}
\label{eq:TImodel2}
\end{equation}
A similar model was proposed in \citet{Larsen1996}:
\begin{equation}
    \frac{\Delta u'_{max}}{U_\infty}= 0.29 \left( x/d\right)^{-1/3}\sqrt{1-\sqrt{1-C_T}}
\label{eq:TImodel3}
\end{equation}

In \citet{Quarton1990}, the downstream evolution of the peak turbulence intensity is modeled by using as scaling parameter the location of the near-wake extent, $x_n$:
\begin{equation}
    \frac{\Delta u'_{max}}{U_\infty}=4.8~C_T^{0.7}I_u^{0.68} \left( x/x_n\right)^{-0.57}
\label{eq:TImodel4}
\end{equation}
where $x_n$ is estimated as \cite{Vermeulen1980}:
\begin{equation}
   x_n=\frac{\sqrt{0.214+0.144m} \left( 1-\sqrt{0.134+0.124m} \right) }{ \left( 1-\sqrt{0.214+0.144m} \right) \sqrt{0.134+0.124m} } \frac{r_0}{dr/dx},
\label{eq:xnVermeulen1980}    
\end{equation}
where $m=1/\sqrt{10C_T}$ and $r_0=0.5d\sqrt{0.5(m+1)}$. The wake growth rate, $dr/dx$, can be estimated as:
\begin{equation}
    \frac{dr}{dx}=\sqrt{\left( \frac{dr}{dx}\right)^2_a+\left( \frac{dr}{dx}\right)^2_m+\left( \frac{dr}{dx}\right)^2_\lambda}
\end{equation}
where the three terms on the right-hand side represent the contribution due to ambient turbulence ($(dr/dx)_a^2=2.5~I_u+0.005$), wake-generated turbulence ($(dr/dx)_m^2=(1-m)\sqrt{1.49+m}/9.76/(1+m)$), and mechanical turbulence ($(dr/dx)_\lambda^2=0.012~N_B~\lambda$, where $N_B$ is the number of blades and $\lambda$ is the rotor tip-speed-ratio).

Another model to estimate the added turbulence intensity was proposed by \citet{Hassan1992}:
\begin{equation}
    \frac{\Delta u'_{max}}{U_\infty}=5.7~C_T^{0.7}I_u^{0.68} \left( x/x_n\right)^{-0.96}
\label{eq:TImodel5}
\end{equation}

For the analysis of the wake-added turbulence intensity for the various clusters, the freestream turbulence intensity, $\sigma_{U_\infty}/U_\infty$, is estimated from the ensemble standard deviation of the LiDAR measurements associated with each cluster, specifically from the median of $\sigma_u$ measured at $x/d=1$ for $|r/d|>0.7$. The respective values, which are reported in Table \ref{tab:Clusters}, show variability among the various clusters similar to that observed for $TI$ estimated from the SCADA data and reported in Table \ref{tab:SCADAmetClusters}. However, the values of $\sigma_{U_\infty}/U_\infty$ are significantly smaller than the respective ones estimated through the nacelle-mounted anemometers, which might be a consequence of the larger measurement volume of the LiDAR and the associated spatial averaging on the velocity fluctuations over each measurement volume \cite{Brugger2016,Puccioni2021}.

\begin{figure}[h!]
    \centering
    \includegraphics[width=\textwidth]{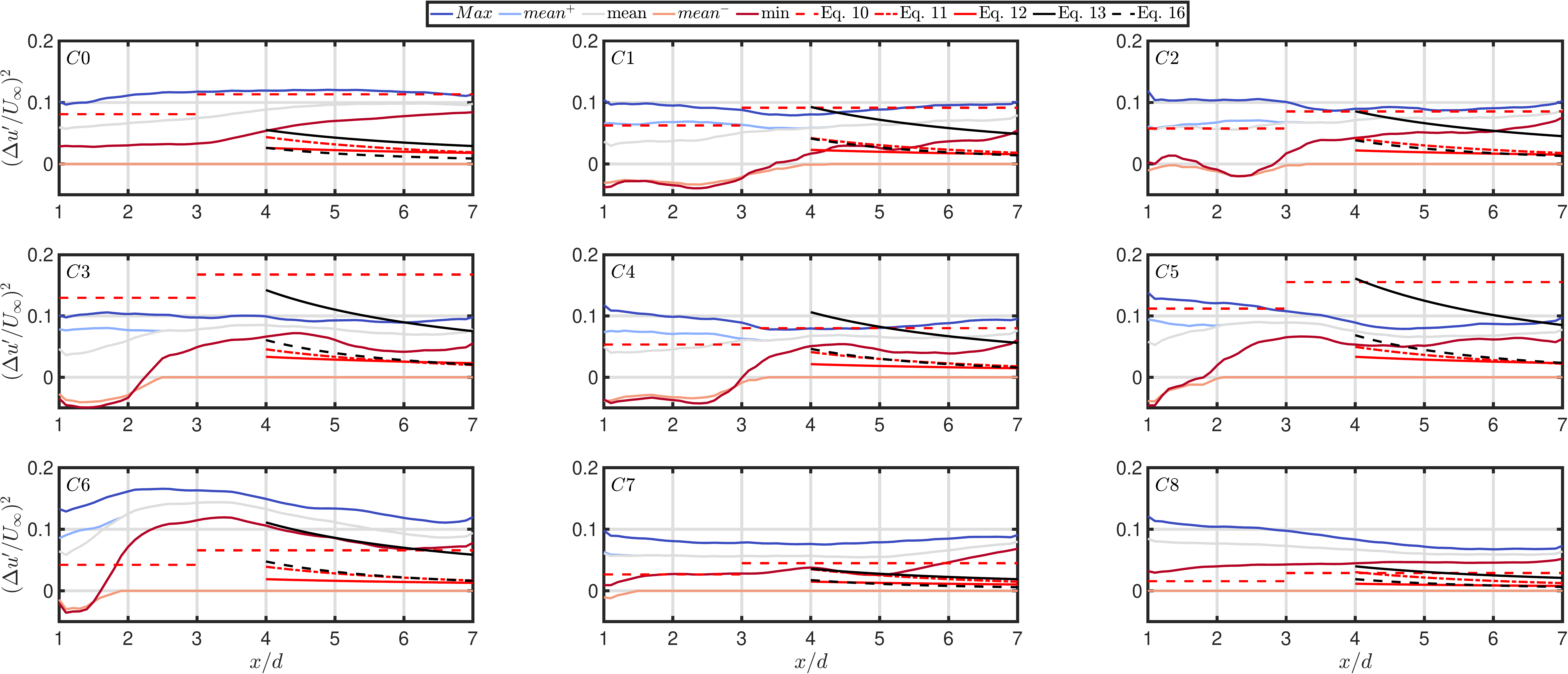}
    \caption{Statistics of the wake-added turbulence intensity, $\Delta u'/U_\infty$, and comparison with the predictions obtained through several engineering models (Eqs. \ref{eq:TImodel1}-\ref{eq:TImodel5}).}
    \label{fig:TIstat}
\end{figure}

We analyze the statistics of the wake-added turbulence intensity in terms of minimum, maximum, and mean of $\Delta u'$; furthermore, mean values for the regions where $\Delta u'$ is only positive or negative are also calculated (referred to as $mean^+$ and $mean^-$, respectively). In Figure \ref{fig:TIstat}, the statistics of $\Delta u'$ indicate that in the near wake, i.e. for $x/d\lessapprox 3$, a region with turbulence intensity smaller than the incoming turbulence intensity ($\Delta u'<0$) is generally observed for the clusters associated with operations in region 2 of the turbine power curve under unstable atmospheric conditions, i.e. for the clusters $C1$-$C6$. This wake feature can be associated with regularization of the flow induced by the wake swirl and the rotational flow induced by the near-wake vorticity structures, such as hub vortex and root vortices present at the core of the wake \cite{Chamorro2009,Iungo2013JFM,Viola2014,Ashton2016}. By proceeding downstream, this region with negative $\Delta u'$ gradually fades out together with the diffusion of the mentioned wake vorticity structures. 

In the near wake, the mean of $\Delta u'$ generally increases as a result of the shear-generated turbulence, which is due to the increasing velocity deficit associated with the thrust force induced by the turbine rotor. Downstream to the near wake, $\Delta u'$ shows a significant reduction only for the clusters $C5$, $C6$, and $C8$. Indeed, the LiDAR velocity statistics in Figure \ref{fig:ClusterStats} show that the magnitude of the $\Delta u'$ peaks located roughly at the lateral boundaries of the wake reduce by proceeding downstream, which is a consequence of the wake recovery, while $\Delta u'$ increases at the wake core, which might be the footprint of far-wake dynamics, e.g. wake meandering \cite{Keck2014, Larsen2015, Muller2015a, Foti2016}.

Predictions of the statistics for the wake-added turbulence intensity obtained through the various above-mentioned models (Eqs. \ref{eq:TImodel1}-\ref{eq:TImodel5}) are also reported in Figure \ref{fig:TIstat}. These models generally produce realistic predictions of the maximum wake-generated turbulence; however, the downstream extent of the wake investigated ($1\leq x/d \leq 7$) is far to show an exponential decay, as usually predicted through these models for the far-wake region. This is a critical limitation of the existing models for predicting the wake-added turbulence intensity because of the lack of accurate predictions for cases where wake interactions may occur with relatively small streamwise spacing among wind turbines, i.e. smaller than 7$d$ \cite{ElAsha2017}. 

\section{Conclusions}
\label{sec:Concl}
LiDAR measurements of wind turbine wakes have been investigated through cluster analysis to identify the most representative wake morphologies associated with different atmospheric stability regimes, wind conditions, and control settings of the wind turbines. The wake LiDAR measurements are first projected on a truncated POD basis consisting of only five physics-informed POD modes. The reduced dimensionality of the experimental dataset has been instrumental to reduce the computational costs for the cluster analysis of two orders of magnitude.

The coefficients of the selected POD modes are then injected in a k-means algorithm, which identifies nine clusters to cover the variability of the wind turbine wakes observed through the LiDAR measurements. The synergistic analysis of the clustered LiDAR, meteorological, and SCADA data has enabled us to ascribe seven clusters to operations in region two of the power curve, namely for incoming wind speeds lower than the turbine rated wind speed, and two clusters for operations above rated wind speed. While the latter mainly occur under neutral/stable atmospheric conditions, the other seven clusters owing to region two of the power curve are characterized by a varying level of atmospheric instability, leading to different incoming turbulence intensity and wake recovery rate.

It is noteworthy that the completely data-driven approach of the cluster analysis, which avoids the imposition of bounds for the various atmospheric, wind, and turbine parameters for detecting the wake variability, allows for the identification of systematic operations of the wind turbines with a certain yaw misalignment from the incoming mean wind direction. Indeed, for a duration of about 10\% of the entire LiDAR experiment, the turbine rotors were positioned with a yaw angle between 3$^\circ$ and 5$^\circ$, leading to significant deflections of the wind turbine wakes, and, eventually, effects on the wind farm power efficiency. 

The clustered LiDAR data have also been analyzed in terms of wake-added turbulence intensity. Specifically, regions with a reduced wind turbulence intensity (negative wake-added turbulence intensity) have been observed in the near wake for operations in region two of the power curve and under unstable atmospheric conditions. This wake feature might be ascribed to the flow re-organization due to the wake vorticity structures, such as tip and hub vortices. In the far-wake, a general decay of the wake-added turbulence intensity is observed, yet not with a systematic exponential decay as proposed from existing wake models. However, predictions of the wake-added turbulence intensity obtained with several existing models have produced similar values to those obtained from the experimental data, albeit their accuracy is relatively poor in terms of streamwise variability. This analysis suggests that further work is needed for modeling the wake-added turbulence intensity for wake regions where the asymptotic exponential decay is not achieved yet, which can be an important flow feature in presence of wake interactions with streamwise spacing smaller than about seven rotor diameters.

\section*{Acknowledgments}
This research has been funded by a grant from the National Science Foundation CBET Fluid Dynamics, award
number 1705837. Pattern Energy Group is acknowledged to provide access to the wind farm for the LiDAR experiment and wind farm data. The Texas Advanced Computing Center is acknowledged for the computational resources. The submitted manuscript has been created by UChicago Argonne, LLC, Operator of Argonne National Laboratory ("Argonne”). Argonne, a U.S. Department of Energy Office of Science laboratory, is operated under Contract No. DE-AC02-06CH11357. This work also utilized the resources of the Argonne Leadership Computing Facility under Contract No. DE-AC02-06CH11357. The U.S. Government retains for itself, and others acting on its behalf, a paid-up nonexclusive, irrevocable worldwide license in said article to reproduce, prepare derivative works, distribute copies to the public, and perform publicly and display publicly, by or on behalf of the Government. The Department of Energy will provide public access to these results of federally sponsored research in accordance with the DOE Public Access Plan (http://energy.gov/downloads/doe-public-access-plan).

\bibliographystyle{unsrtnat}
\bibliography{Wakes}

\end{document}